\documentclass[12pt]{article}
\usepackage[utf8]{inputenc}
\usepackage{amsfonts,epsfig}
\usepackage[hyphens]{url}
\usepackage[hidelinks]{hyperref}
\usepackage{breakurl}
\usepackage[authoryear]{natbib}

\usepackage{geometry} 
\usepackage{sectsty} 
\usepackage[normalem]{ulem} 
\sectionfont{\sffamily\bfseries\upshape\large}
\subsectionfont{\sffamily\bfseries\upshape\normalsize} 
\subsubsectionfont{\sffamily\bfseries\upshape\normalsize} 
\makeatletter

\usepackage{subcaption}
\usepackage{graphicx}
\usepackage{wrapfig}
 \usepackage{graphicx,fullpage,amsmath,multicol,multirow,amssymb,amsbsy,pifont}
 \usepackage{setspace}
 \usepackage{epsfig}
 \usepackage{amsmath, amsthm, amssymb, bm}
 \usepackage{color}
\usepackage{graphics}
\usepackage{booktabs}
\usepackage{makecell}
\usepackage{caption,array}

\renewcommand\@seccntformat[1]{\csname the#1\endcsname.\quad}

\numberwithin{figure}{section}
\numberwithin{table}{section}
\numberwithin{equation}{subsection} 
\makeatletter
\@addtoreset{equation}{section}
\makeatother
\makeatletter
\def\@maketitle{%
 \begin{center}%
 \let \footnote \thanks
  {\Large \@title \par}%
  {\normalsize
   \begin{tabular}[t]{c}%
    \@author
   \end{tabular}\par}%
  {\small \@date}%
 \end{center}%
}
\makeatother

\newcommand{\btR}{\vspace{-.16in}\begin{quotation}\begin{small}\noindent\begin{verbatim}}

\title{\bf Multilevel Regression and Poststratification Interface: An Application to Track Community-level COVID-19 Viral Transmission\footnote{The interface is available on CRAN: \url{https://cran.r-project.org/web/packages/shinymrp/index.html}.} \vspace{.1in} }
\author{Yajuan Si,\footnote{University of Michigan, Ann Arbor.} \,
Toan Tran,$^{\dagger}$\,
Jonah Gabry,\footnote{Columbia University, New York} \, 
 Mitzi Morris,$^{\ddagger}$\,
 and Andrew Gelman$^{\ddagger}$
  \vspace{.1in}}
\date{\vspace{-.1in}}
\begin{document}
\maketitle
\thispagestyle{empty}

\begin{abstract}

We present a novel Bayesian workflow for multilevel regression and poststratification (MRP), introducing extensions to time-varying data and granular geography and publicly available open-source computation tools, facilitating broad research adoption and reproducibility. In the absence of comprehensive or random testing throughout the COVID-19 pandemic, we have developed a proxy method for synthetic random sampling to estimate community-level viral incidence, based on viral RNA testing of asymptomatic patients who present for elective procedures within a hospital system. The approach collects routine testing data on SARS-CoV-2 exposure among outpatients and performs statistical adjustments of sample representation using MRP, a procedure that adjusts for nonrepresentativeness of the sample and yields stable small group estimates. We illustrate the MRP interface with an application to track community-level COVID-19 viral transmission in the state of Michigan.

\noindent{\em \bf Keywords:} Bayesian workflow, bias correction, subgroup estimation, surveillance system, reproducibility

\end{abstract}

\section{Introduction}

Early and accurate knowledge of incidence and trends of transmission within communities is crucial for monitoring a pandemic and supporting policymakers in assessing the effects of restrictive measures on individual and community behaviors. However, without universal screening or random testing, government policy and healthcare implementation responses have relied on testing people who were symptomatic or presumed exposed, with policies guided by officially reported positivity rates and counts of positive cases in the community. These data are biased in concept and flawed in practice. It is essential to establish an operational surveillance system that allows prompt assessments of mitigation efforts and future predictions of clinical burdens. This would trigger an effective healthcare response and inform other epidemics. Since universal random testing is not always feasible, we must develop alternative methods that offer similar advantages. An effective proxy measure should be able to detect increases in viral incidence before these trends become clinically relevant. It should also identify decreases in incidence, allowing for the safe suspension of mitigation strategies and supporting economic and social recovery. Ideally, the data collection and analysis procedure would be practical at the community level and applicable nationwide, with the capacity to focus on burdens within specific demographics. It must also be reliable, statistically valid, cost-effective, and automatic, ensuring that it can be deployed promptly in future pandemics.

%In this paper, we present a user-friendly interface with an automatic implementation of 
We present a novel Bayesian workflow for multilevel regression and poststratification \citep[MRP,][]{gelman:little:97}, which is built on a user-friendly open-source interface, and demonstrate the application to track community-level COVID-19 viral transmission. Previous work has developed the foundation of such a proxy metric for COVID-19 tracking data collection and statistical adjustment of demographic representation~\citep{mrp-covid21,mrp-covid22}. The approach collects electronic health records (EHRs) on routine viral testing of patients who present for elective procedures within a hospital system and performs MRP to estimate actual viral trends. The findings in a diverse urban-suburban-rural setting in Indiana show that this model predicts the clinical burden of SARS-CoV-2 earlier and more accurately than the currently accepted metrics. In contrast, the official testing data fail to inform the surge of clinical burdens. 

Our new development enhances traditional MRP by introducing two novel methodological extensions: estimation varying across time and at granular geographic levels. Current applications of MRP predominantly rely on cross-sectional data collected at a single time point, either through probability sample surveys \citep[e.g.,][]{Zhang15-mrp,prior-si2018} or nonprobability samples \citep[e.g.,][]{wang:gelman14, mrp-si20}. In contrast, this paper extends MRP to accommodate time-varying data, enabling the tracking of trends. Moreover, due to sample size constraints, existing MRP estimates are typically at the national or state level. Our interface overcomes these limitations by supporting analyses at multiple geographic resolutions, including ZIP codes, counties, and states. The interface is capable of modeling both binary and continuous outcomes. As a demonstration, we apply the interface to time-varying, ZIP-code-level data to monitor demographic and county-level COVID-19 viral transmission trends in Michigan. We generate granular, timely and reproducible findings to direct resource allocation, guide surge prediction and inform rapid public health responses and evidence-based policy making.

As a prediction approach to modeling the outcome measures with individual-level and group-level predictors, MRP has become increasingly popular for subgroup estimation. Originally applied to estimate state-level public opinions from sociodemographic subgroups using sample surveys, MRP has two key components: (1) multilevel regression for small group estimation by setting up a predictive model with a large number of covariates and regularizing with Bayesian prior specifications, and (2) poststratification to adjust for selection bias and correct for imbalances in the sample composition from the target population. Flexible modeling of study outcomes can capture complex data structures conditional on poststratification cells, which are determined by the cross-tabulation of categorical auxiliary variables and calibrate the sample discrepancy with population control information. Besides applications in social sciences, especially in election forecasting \citep[e.g.,][]{wang:gelman14,Yougov:mrp20}, MRP has also shown promise in public health research \citep[e.g.,][]{Zhang15-mrp,Carlin:AJE2018,carlin:AJE20,prior-si2018}. \cite{mrp-si20} show that the key to the success of MRP in applications is the inclusion of highly predictive covariates, and \cite{emrp21} discuss estimation approaches when the population distribution of the poststratification variables is unknown. 

The new workflow improves the MRP method with an application of tracking community-level transmission---across geographic areas and demographic subgroups---to monitor the epidemic over time using the developed interface as an operational surveillance system. Leveraging the statistical programming language Stan~\citep{stan-software:2021} to conduct Bayesian computation and model estimation, specifically with the R package \texttt{cmdstanr}~\citep{cmdstanr}, the user-friendly interface promotes methodological transparency, reproducibility, and standardization for public health and social science research. The interface is accessible via both web-based and local platforms using the R package \texttt{shinymrp}~\citep{shinymrp-cran}, offering flexible tools to facilitate analyses for diverse users. By integrating data preparation, model fitting, diagnostic checks, and graphical reporting into a comprehensive end-to-end pipeline, the interface enables researchers and practitioners to standardize MRP analyses across studies and datasets. This feature substantially lowers technical barriers, allowing reproducible application of advanced statistical methods and fostering comparability in population health research. We illustrate model use and interface capabilities with a real-world application tracking community-level COVID-19 viral transmission in Michigan. Health agencies can use subgroup-specific incidence estimates to allocate testing resources and direct targeted interventions to communities with greater vulnerability. Tracking epidemiological trends in real time through routine hospital data provides actionable guidance for surge prediction and rapid public health responses. While our primary application centers on community-level COVID-19 monitoring, the MRP workflow developed here is broadly applicable across a range of public health surveillance contexts.

The main contributions of this paper include: 1) introducing the MRP computational interface; 2) extending MRP models with time-varying and granular geographic data; 3) applying the metric to track COVID-19 viral transmission in Michigan. We describe the data source and MRP methods in Section~\ref{method}. Section~\ref{workflow} presents the workflow of statistical analyses, from data preprocessing, descriptive summaries, model fitting and diagnostics, to result presentation and validation. Section~\ref{discussion} concludes with discussions and potential directions for future work. 

\section{Methods}
\label{method}

\subsection{Data}
\label{data}

Our COVID-19 tracking approach collects electronic health records of prospective surgical (and other invasive procedure) patients who are asymptomatic and have tested for acute SARS-CoV-2 infection before performing the procedure \citep{mrp-covid21, mrp-covid22}. Upon the reopening of elective medical and surgical procedures after the initial COVID-19 outbreak in early 2020, all preoperative patients were uniformly required---per the American Society of Anesthesiology guidelines---to undergo surgical risk evaluation and testing for acute SARS-CoV-2 infection before any such procedures. This policy was implemented nationwide across the U.S. All elective patients were presumed asymptomatic because any individual reporting symptoms or recent exposure to the virus would have their procedure either canceled or deferred. Using a standardized protocol, all preoperative patients underwent polymerase chain reaction (PCR) testing for viral RNA four days prior to their scheduled procedure, with tests administered by health system staff and samples analyzed using the same system. This PCR testing protocol was maintained consistently throughout the study period. Additionally, a subset of patients, for whom preoperative blood testing was clinically indicated based on age, health status, or surgery type, were also screened for the presence of immunoglobulin G (IgG) to the SARS-CoV-2 nucleocapsid protein (IgG N), beginning May 1, 2020.

In collaboration with hospital database managers and in compliance with HIPAA privacy regulations, we collected EHR data including PCR test results, test dates, sex, age, race, and five-digit ZIP codes. The group represents broad age, racial, and socioeconomic diversity, with its only explicit correlation to disease status being the selection for elective surgical procedures and absence of symptoms or known exposure. We assume that, within any demographic and geographic stratum, the ratio of asymptomatic to symptomatic SARS-CoV-2 infections remains constant. Accordingly, the incidence of asymptomatic infections should proportionally reflect community-wide viral incidence and can serve as a proxy for true incidence trends, though this ratio may change with the emergence of new viral variants and the level of acquired immunity over time. To the extent that healthcare use or other factors affect the selection and ratio, we expect much of this variation to be addressed through our model adjustments. We discuss the potential violation of these assumptions in Section~\ref{discussion}. MRP adjusts the demographic (sex, age, and race) and geographic (five-digit ZIP code) distributions to the target population. The target population is defined as U.S. residents dwelling in the catchment area of the collected ZIP codes. We link the input patient EHR data with ZIP codes to census tract measures in the American Community Survey (ACS), the largest household survey of the U.S. population \citep{acs2021}, and use the ACS aggregated summaries of sociodemographic and socioeconomic characteristics as geographic predictors at the ZIP level.

Previous work has treated PCR test sensitivity and specificity as unknown parameters, incorporating information from prior studies and accounting for estimation uncertainty in final MRP estimates~\citep{specificity:gelman20,mrp-covid21,mrp-covid22}, which we recommend for use when their values are undetermined. The interface allows for specifying different sensitivity and specificity values; for our demonstration, we presume 70\% clinical sensitivity and 100\% specificity, consistent with the previous setting during the same study period (March 2020--October 2022)~\citep{bendavid2021covid,specificity:gelman20,mrp-covid21,mrp-covid22}.

We track viral infections on a weekly basis. Below, we first introduce the conventional MRP framework for cross-sectional data, followed by extensions to accommodate time-varying data features. 
 
\subsection{MRP for cross-sectional data}

MRP first fits a multilevel regression model to predict the outcome measure as a function of a set of factors, then poststratifies the categorical factors so that their distributions match those of the target population. We use a binary outcome of interest as an example. Let $y_i (=0/1)$ be the binary response for individual $i$, with $y_i=1$ indicating the positive response. We employ a logistic regression with varying effects for age, race, and ZIP code, where the ZIP-code-level variation is further explained by the ZIP-code-level predictors.
\begin{align}
\label{mrp-1}
\textrm{Pr}(y_i = 1) = \textrm{logit}^{-1}(
\beta_1+\beta_2{\rm male}_i +
\alpha_{\rm a[i]}^{\rm age}
+ \alpha_{\rm r[i]}^{\rm race}
+ \alpha_{\rm s[i]}^{\rm ZIP}
),
\end{align}
where ${\rm male}_i$ is an indicator for men, $\alpha_{\rm a}^{\rm age}$ is the age effect, with a value of $a[i]$ for subject $i$, on the log-odds function of the probability of having a positive response, $\alpha_{\rm r}^{\rm race}$ is the racial effect, and $\alpha_{\rm s}^{\rm ZIP}$ is the ZIP-code-level effect. In the Bayesian framework, we assign hierarchical priors to varying intercepts as a default setting:
\begin{align}
\label{prior}
\nonumber &\alpha^{\rm age} \sim \mbox{normal}(0,\sigma^{\rm age} ), \,\,\, \sigma^{\rm age}\sim \mbox{normal}_+ (0,2.5)\\
&\alpha^{\rm race} \sim \mbox{normal}(0,\sigma^{\rm race} ), \,\,\, \sigma^{\rm race}\sim \mbox{normal}_+ (0,2.5).
\end{align}
Here $\mbox{normal}_+ (0,2.5)$ represents a half-normal distribution with the mean $0$ and standard deviation $2.5$ restricted to positive values. As we have ZIP-code-level predictors $\vec{Z}^{\rm ZIP}_{s}$, we need to build another model in which $\alpha_{\rm s}^{\rm ZIP}$ is the outcome of a linear regression with ZIP-code-level predictors:
\begin{align}
\label{prior-zip}
\alpha_{\rm s}^{\rm ZIP} =\vec{\alpha}\vec{Z}^{\rm ZIP}_{s} +  e_s, \,\,\, e_s\sim \mbox{normal}(0,\sigma^{\rm ZIP} ),\,\,\, \sigma^{\rm ZIP}\sim \mbox{normal}_+ (0,2.5),
\end{align}
where $e_s$ is a ZIP-code-level random error.

Alternative prior distributions, including structured priors for high-order interaction terms~\citep{prior-si2018} and spatial prior accounting for geospatial correlation, can also be specified. We use the default setting in the interface with normal priors given in \eqref{prior} and \eqref{prior-zip} as examples and discuss extensions to other settings in Section~\ref{discussion}.

Because \eqref{mrp-1} assumes that the people in the same poststratification cell share the same response probability, we can replace the microdata with cellwise aggregates and employ a binomial model for the sum of the responses in cell $j$ as $y^*_j \sim \textrm{binomial}(n_j, \theta_j)$, where $n_j$ is the sample cell size and $\theta_j=\textrm{logit}^{-1}(
\beta_1+\beta_2{\rm male}_j +
\alpha_{\rm a[j]}^{\rm age}
+ \alpha_{\rm r[j]}^{\rm race}
+ \alpha_{\rm s[j]}^{\rm ZIP}
)
$ using the cellwise effects of all factors. The input data for analysis can thus be either microdata or cellwise aggregates. 

To generate overall population or subgroup estimates, we combine model predictions within the poststratification cells---in the contingency table of sex, age, race, and ZIP---weighted by the population cell frequencies $N_j$, which are derived from the linked ACS data in our application. Additionally, we may choose custom poststratification data for specific target populations (e.g., a different country, rather than the U.S.). If we write the expected outcome in cell $j$  based on model~\eqref{mrp-1} as $\hat{\theta}_j$ in cell $j$, the population average from MRP is then:
$$
\hat{\theta}^{\rm pop} = \frac{\sum_j N_j \hat{\theta}_j}{\sum_j N_j}.
$$
The MRP estimator for county $c$ aggregates over covered cells $j$ in that county as,
$$
\hat{\theta}_s^{\rm pop} = \frac{\sum_{j \in \textrm{county c}} N_j \hat{\theta}_j}{\sum_{j \in \textrm{county c}} N_j}.
$$
We implement Bayesian inference for the estimates, where the variance estimates and 95\% credible intervals are computed based on the posterior samples.

\subsection{MRP for time-varying data}

As an example of time-varying data, we model weekly PCR testing results. Here, MRP proceeds in two steps: (1) fit a multilevel model to the testing data for incidence incorporating time and covariates, and (2) poststratify using the population distribution of the adjustment variables: sex, age, race, and ZIP codes, where we assume the population distribution is the same during the study period. Hence, the poststratification cell is defined by the cross-tabulation of sex, age, race, ZIP code, and indicators of time in weeks based on the test result dates.

We denote the PCR test result for individual $i$ as $y_i$, where $y_i=1$ indicates a positive result and $y_i=0$ indicates negative. Similarly, with poststratification cells, we assume that people in the same cell have the same infection rate and can directly model cellwise summaries. We obtain aggregated counts as the number of tests $n_j$ and the number of positive cases $y^*_j$ in cell $j$. Let $p_j=\textrm{Pr}(y_{j[i]}=1)$ be the probability that person $i$ in cell $j$ tests positive. We account for the PCR testing sensitivity and specificity, where the positivity $p_j$ is a function of the test sensitivity $\delta$, specificity $\gamma$, and the true incidence $\pi_j$ for people in cell $j$:  
\begin{align}
\label{positivity}
p_j=(1-\gamma)(1-\pi_j )+\delta \pi_j.
\end{align}

We fit a binomial model for $y^*_j$, $y^*_j \sim \textrm{binomial}(n_j, p_j)$ with a logistic regression for $\pi_j$ with covariates---sex, age, race, ZIP codes, and time in weeks---to allow time-varying incidence in the multilevel model.
\begin{align}
\label{pi}
\textrm{logit}(\pi_j)=\beta_1+\beta_2{\rm male}_j+\alpha_{{\rm a}[j]}^{\rm age}+\alpha_{{\rm r}[j]}^{\rm race}+\alpha_{{\rm s}[j]}^{\rm ZIP}+\alpha_{{\rm t}[j]}^{\rm time},
\end{align}
where ${\rm male}_j$ is an indicator for men; ${\rm a}[j]$, ${\rm r}[j]$, and ${\rm s}[j]$ represent age, race, and ZIP levels; and ${\rm t}[j]$ denotes the time in weeks when the test result is collected for cell $j$. We include ZIP-code-level predictors $\vec{Z}^{\rm ZIP}_{s}$ for ZIP code $s$,
\[
\alpha_{s}^{\rm ZIP} =\vec{\alpha}\vec{Z}^{\rm ZIP}_{s} +  e_s.
\]
We assign the same priors in \eqref{prior} and \eqref{prior-zip} to varying intercepts and error terms $e_s$. As to time-varying effects, we assume $\alpha_{{\rm t}}^{\rm time} \sim \mbox{normal}(0,\sigma^{\rm time} )$, with a weakly informative hyperprior, $\sigma^{\rm time}\sim \mbox{normal}_+ (0,5)$.

As an example, we assign normal priors to the ZIP-code-level and time-varying effects. The interface leverages Stan’s modeling capabilities to allow alternative prior choices and can be extended with advanced modeling~\citep{BNFP:SI15}, such as spatial priors for ZIP-code-level effects or time series priors (e.g., first-order autoregressive) for temporal effects. Alternative outcome models (e.g., negative binomial) can be specified to accommodate overdispersion. In our COVID-19 application, we did not find substantial differences in the examined group estimates with various outcome model and prior specifications, so we presented the results based on a binomial model with normal priors. We elaborate further on model extensions in Section~\ref{discussion}.

Using the estimated incidence $\hat{\pi}_j$ based on the Bayesian model in~\eqref{pi}, we adjust for selection bias by applying the sociodemographic distributions in the community with population cell counts $N_j$ based on the ACS, yielding population-level weekly incidence estimates:
\[
\hat{\pi}_{t} = \frac{\sum_{j \in \mbox{week\,} t} N_j\hat{\pi}_j}{\sum_{j \in \mbox{week\,} t} N_j}, 
\]
which can be restricted to specific subgroups or regions of interest, as another key property of MRP is to yield robust estimates for small groups. We obtain the Bayesian credible intervals from the posterior samples for inference.

\section{Bayesian workflow with MRP}
\label{workflow}

The interface implements an end-to-end Bayesian MRP workflow of statistical analyses, from data preprocessing, descriptive summaries, model fitting, diagnostics, to presentation of results, following the principles of \cite{workflow:gelman21}. For illustration, we apply this process to COVID-19 tracking in Michigan and validate the findings in comparison with other studies.

\subsection{Data preprocessing}

\begin{figure}[htp]
\begin{tabular}{c}
\includegraphics[width=0.95\textwidth]{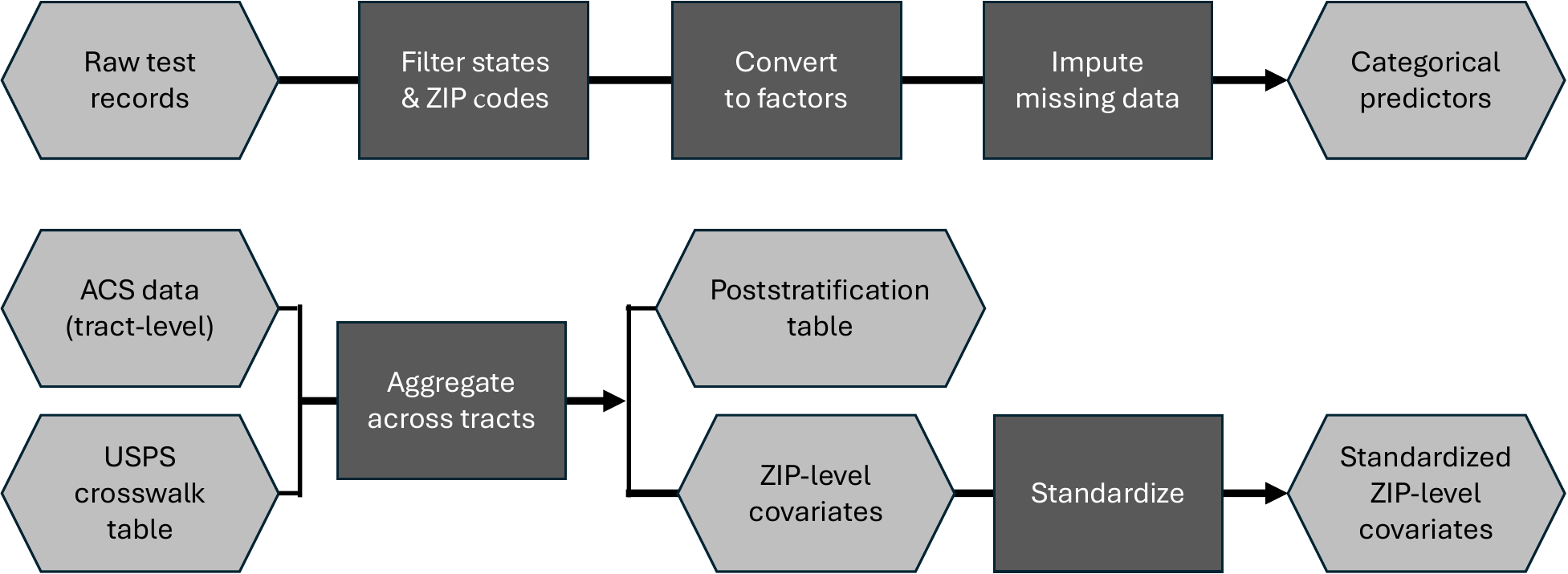}
\end{tabular}
\caption{Data preprocessing flowchart in the interface.}
\label{clean}
\end{figure}

The input data can be either individual patient test records or aggregated summaries at the poststratification cell level. The data cleaning and linking process is illustrated by the flowchart in Figure~\ref{clean}. This workflow automatically imputes missing predictor values using observed frequency distributions, converts categorical variables to factors, and standardizes continuous predictors at the ZIP code level. The MRP integrates three data sources: (1) PCR test results, (2) poststratification cell counts from the ACS, and (3) ZIP-code-level predictors linked from the ACS.

The PCR test records from hospitals include key demographic, geographic, and temporal measures: sex, race, age, five-digit ZIP code, PCR test result, and result date, for the modeling. Balancing operational feasibility, timeliness and accuracy, patient records are linked to the 2021 five-year ACS dataset by residential ZIP codes, using the R package \texttt{tidycensus}~\citep{walker2021package}. This linkage serves two purposes: (1) defining the target population as people living in the ZIP codes’ catchment area and deriving population counts for poststratification cells, and (2) incorporating area-level predictors of viral infection to adjust for geographic variation. While the ACS reports geography at the levels of census tracts, counties, and states, ZIP codes are defined by the U.S. Postal Service (USPS). We use the ZIP code crosswalk table released by the U.S. Department of Housing and Urban Development and USPS to link ZIP codes to census tracts~\citep{hud_usps_ziptract_crosswalk_2022} and calculate the ZIP-code-level measures by aggregating all available tract-level measures weighted by tract population counts. We select the county with the most-overlapping residential addresses for one ZIP code as the ZIP-linked county. The catchment area covered by the list of residence address ZIP codes provided by the Michigan Medicine patients can cover multiple states, beyond Michigan. We filter the data geographically by first removing ZIP codes with five or fewer records and then states that constitute less than 1\% of the remaining data. We construct poststratification cells by cross-classifying sex, race, age, and ZIP code and obtain the population counts for these cells from weighted ACS sample distributions in the relevant catchment area. These counts are assumed to remain constant throughout the study period (2020--2022).

The geographic predictors include both individual-level variables (such as education, employment, and income) and tract-level variables (including urbanicity and the Area Deprivation Index  \citep[ADI,][]{kind2018making}. These are aggregated to ZIP codes as follows: (1) urbanicity: the percentage of covered census tracts classified as urban, weighted by tract population; (2) college: the percentage of residents with an Associate’s degree or higher; (3) poverty: the percentage of residents with incomes below the poverty level in the past year; (4) employment: the percentage of the civilian labor force that is employed; (5) income: the population-weighted average of tract-level median household incomes over the past 12 months; and (6) ADI: the population-weighted average of tract-level ADI values across covered census tracts.
  
\subsection{Descriptive statistics}

We examine descriptive statistics of observed positivity across time and counties, demographics based on individual records, and characteristics of the covered geographic areas. The observed viral infection shows variation across time, geography, and demographic groups. 

\begin{figure}[htp]
\begin{tabular}{cc}
\includegraphics[width=0.47\textwidth]{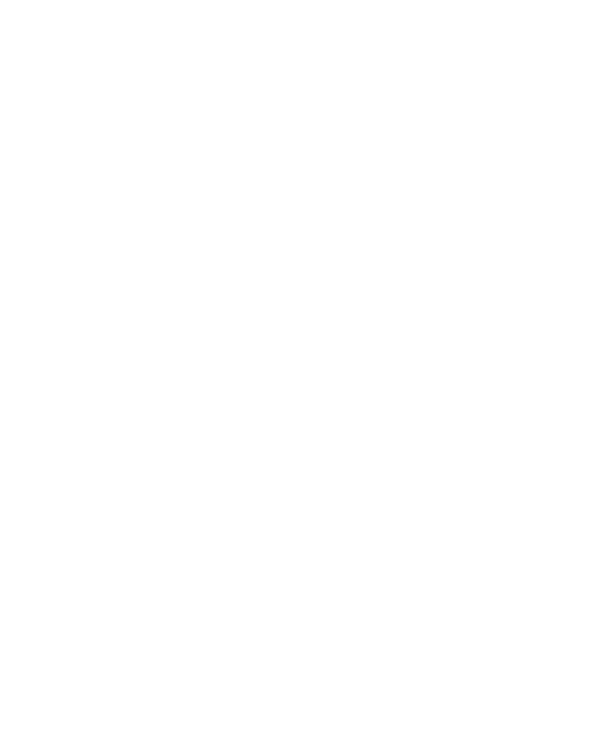} & 
\includegraphics[width=0.47\textwidth]{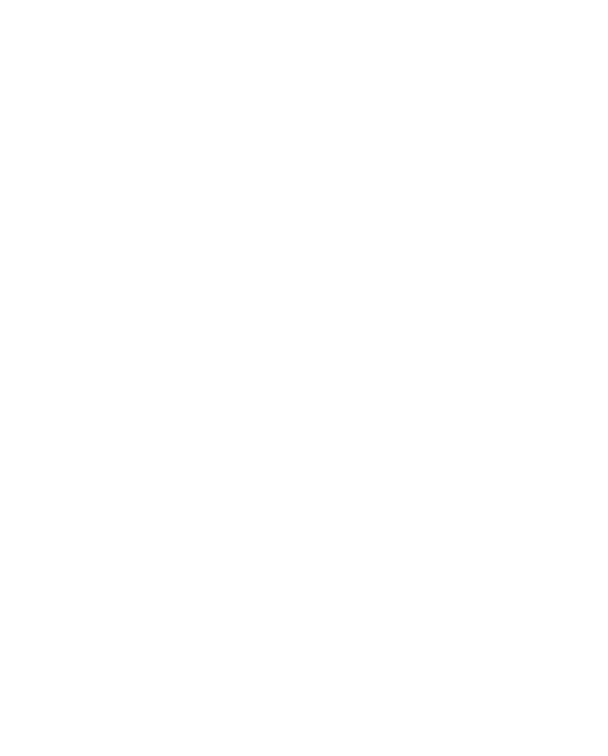} 
\end{tabular}
\caption{Highest values among weekly positive response rates (left) and available sample sizes (right) across 94 counties in the catchment area, exemplifying the large geographic variation with sparse data.}
\label{county}
\end{figure}

Figure~\ref{county} presents the highest value among weekly positivities and collected sample sizes across counties, exemplifying the large geographic variation. Most people are from the four counties in Southeast Michigan, where the medical center is located. However, the sample catchment area covers 94 counties. The test positivity among asymptomatic patients is often lower than 1\%, but greater variability in counties with a small number of tests results in higher than 80\% positivity in some cases. The geographically adjacent areas may not share similar peak values. 
  
\begin{figure}[htp]
\begin{tabular}{lll}
\includegraphics[width=0.32\textwidth, height=2.5in]{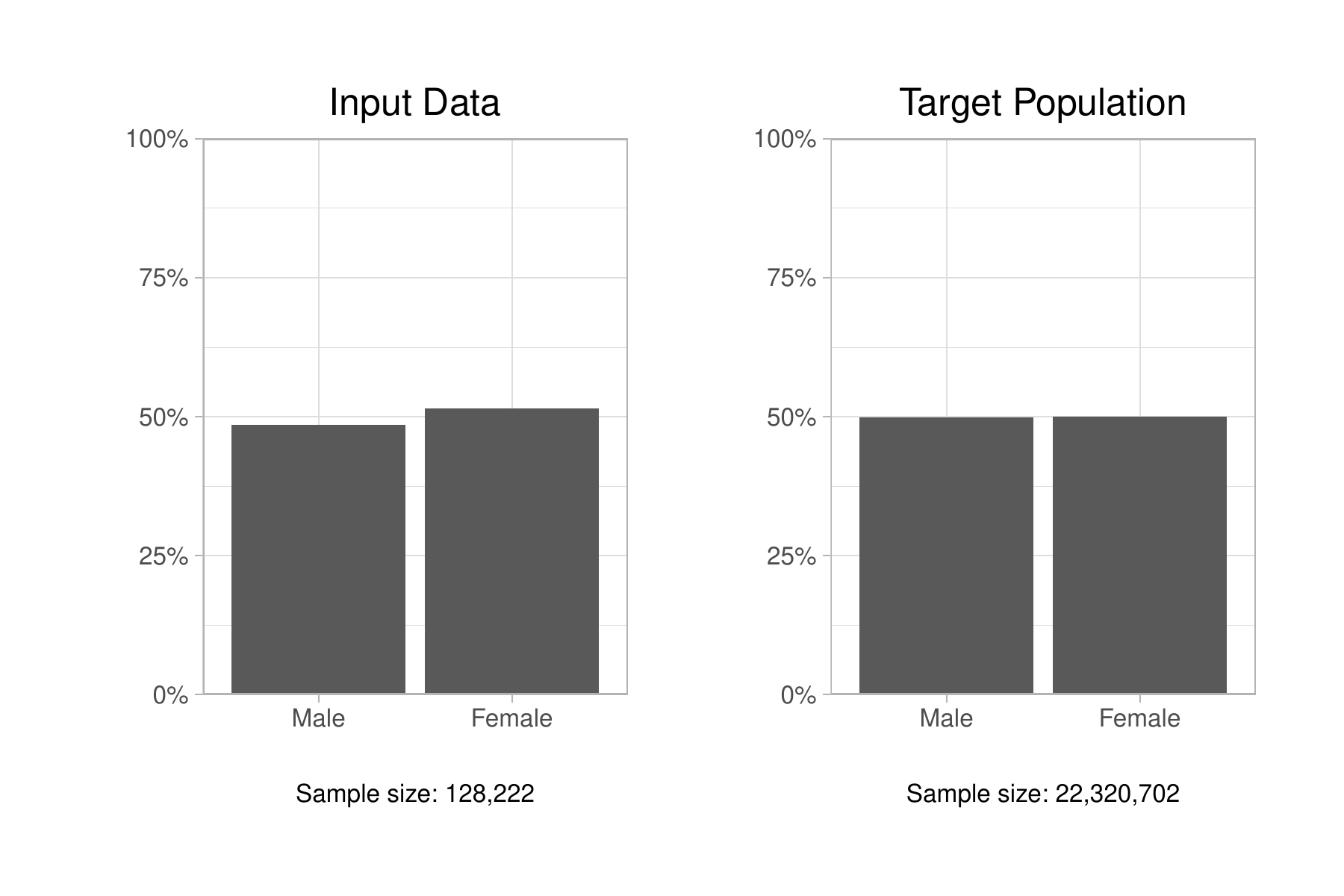} & 
\includegraphics[width=0.32\textwidth, height=2.5in]{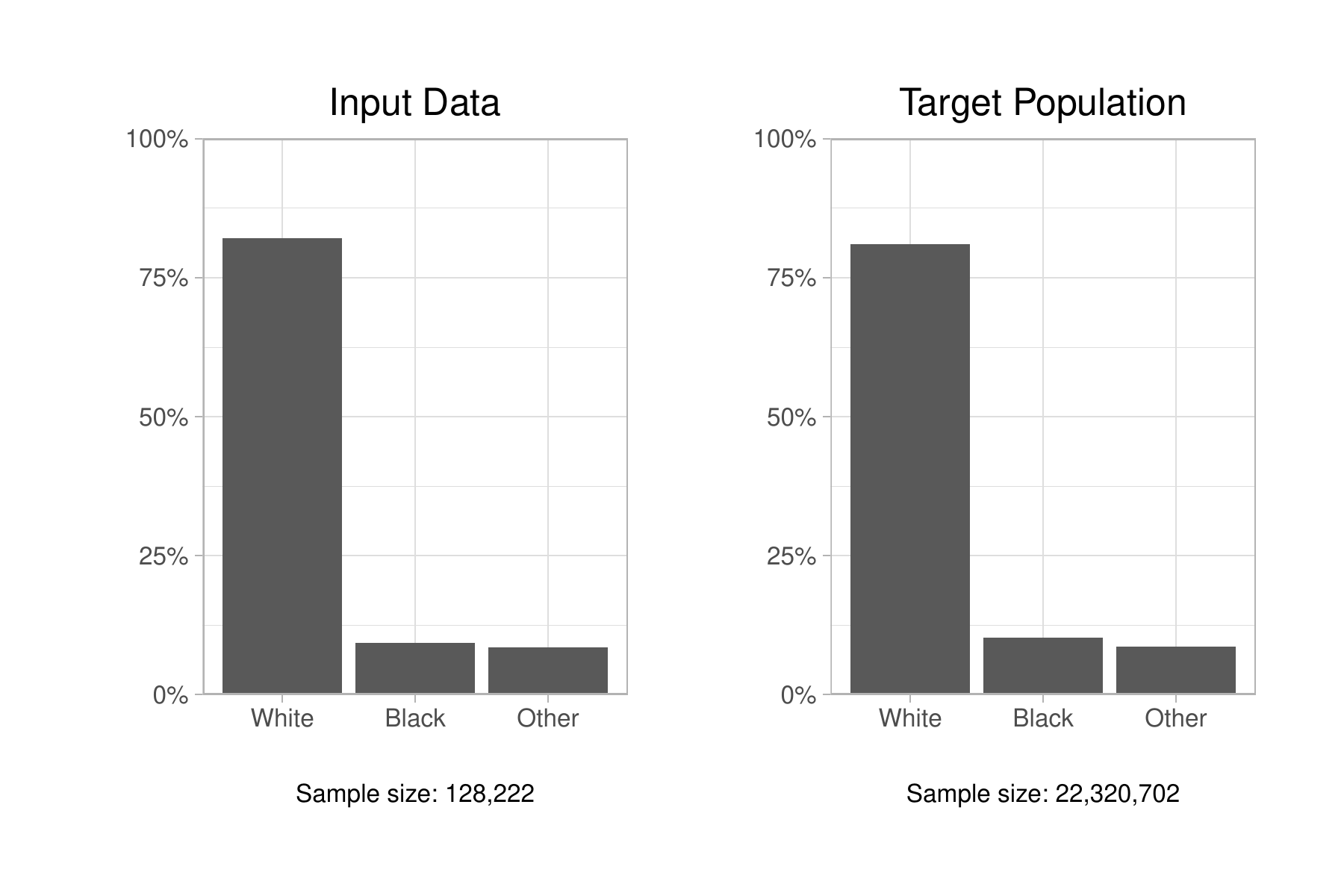} &
\includegraphics[width=0.32\textwidth, height=2.5in]{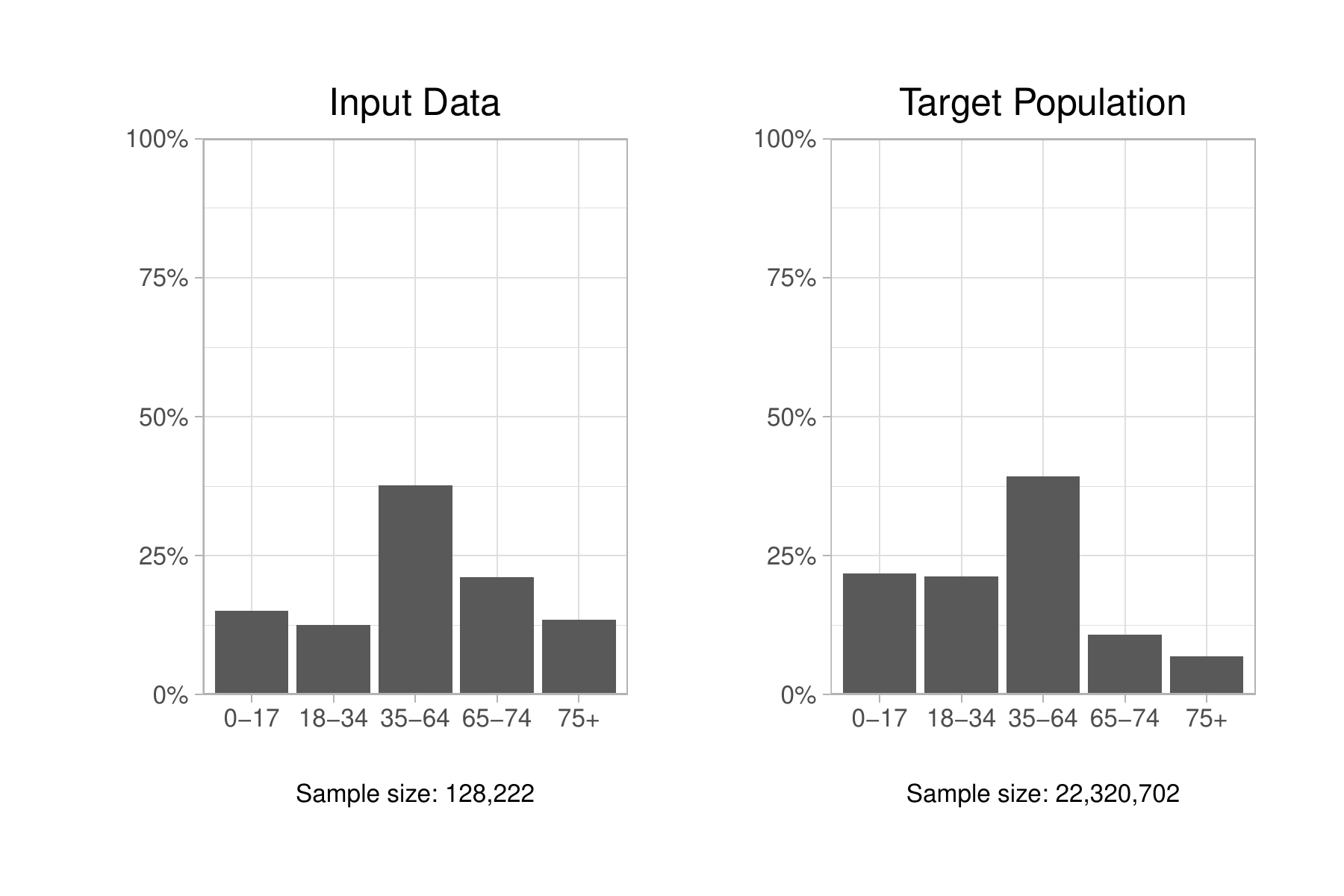} 
\end{tabular}
\caption{Comparisons of demographic distributions between the input data of hospital patients and the target population approximated by linked American Community Survey data in the catchment area. The hospital patients over-represent female, White, and older people.}
\label{dem}
\end{figure}

Figure \ref{dem} compares the sex, race, and age distributions between the hospital patients ($n=128,\!222$) and the population residing in the catchment area ($N=22,\!320,\!702$). The hospital patients have larger proportions of female, White, and older people than the population, and this sample discrepancy will be adjusted by the poststratification step in MRP. 

% \begin{figure}[htp]
% \begin{tabular}{cc}
% \includegraphics[width=0.475\textwidth, height=2.9in]{plot/urbanicity_geo.pdf} & 
% \includegraphics[width=0.475\textwidth, height=2.9in]{plot/adi_geo.pdf} \\
% \includegraphics[width=0.475\textwidth, height=2.9in]{plot/college_geo.pdf} & 
% \includegraphics[width=0.475\textwidth, height=2.9in]{plot/employment_geo.pdf} \\
% \includegraphics[width=0.475\textwidth, height=2.9in]{plot/income_geo.pdf} & 
% \includegraphics[width=0.475\textwidth, height=2.9in]{plot/poverty_geo.pdf} 
% \end{tabular}
% \caption{Distributions of geographic characteristics based on the linked American Community Survey data in the catchment area.}
% \label{ZIP}
% \end{figure}

Figure \ref{ZIP} in the Appendix presents the distributions of geographic characteristics. The catchment area of the hospital patients' residence covers 848 ZIP codes and has a broad and diverse representation in terms of urban/rural areas, area deprivation status, higher education attainment, employment rate, income, and poverty. We expect that socioeconomic measures at the ZIP level would affect individual behaviors and health and be related to viral transmission. The geographic characteristics would explain the spatial variation. The poststratification uses the population counts by ZIP code but does not adjust the geographic characteristics. 

\subsection{Model fitting}

We fit the model via Bayesian computation with Markov chain Monte Carlo in Stan and obtain model fit summaries and convergence assessments of the posterior sample of model parameters. We give an example of output from model \eqref{pi} in Section~\ref{fit_print} of the Appendix. With the interface, we can also specify and fit different models with various choices of individual/geographic covariates and varying/non-varying effects under different prior distributions.

\subsection{Model diagnostics}
We compare different models and present model diagnostic results. We employ the approximate leave-one-out cross-validation (LOO-CV) implemented in the R package \texttt{loo}~\citep{loo-R} and posterior predictive checking \cite[PPC;][]{ppc:gelman-96}. The LOO-CV assesses the posterior predictive performance of Bayesian models and compares different models on expected log predictive density (elpd) for new data.

We have compared the three models with different mean structure and variance specifications, with included predictors listed in Table~\ref{model:display}. Model A includes coefficients for sex and geographic predictors and varying effects of age, race, time in weeks, and ZIP code. Model B adds high-order interactions, between race and college attainment status, to Model A. Model C removes the ZIP-varying effects from Model A. 

% {\tiny
% \begin{align*}
%   \textrm{Model A: } &\beta_1+\beta_2{\rm male}_j+\alpha_{{\rm a}[j]}^{\rm age}+\alpha_{{\rm r}[j]}^{\rm race}+\alpha_{{\rm t}[j]}^{\rm time}+\alpha_1{\rm ADI}_{{\rm s}[j]} +\alpha_2{\rm college}_{{\rm s}[j]}+\alpha_3{\rm employment}_{{\rm s}[j]} +\alpha_4{\rm income}_{{\rm s}[j]}+\alpha_5{\rm poverty}_{{\rm s}[j]} +\alpha_6{\rm urbanicity}_{{\rm s}[j]} + e_{{\rm s}[j]}\\
%   \textrm{Model B: } &\beta_1+\beta_2{\rm male}_j+\alpha_{{\rm a}[j]}^{\rm age}+\alpha_{{\rm r}[j]}^{\rm race}+\alpha_{{\rm t}[j]}^{\rm time}+\alpha_1{\rm ADI}_{{\rm s}[j]} +\alpha_2{\rm college}_{{\rm s}[j]}+\alpha_3{\rm employment}_{{\rm s}[j]} +\alpha_4{\rm income}_{{\rm s}[j]}+\alpha_5{\rm poverty}_{{\rm s}[j]} +\alpha_6{\rm urbanicity}_{{\rm s}[j]}\\
% &+\alpha_7{\rm race * college}_{{\rm s}[j]} + e_{{\rm s}[j]}\\
% \textrm{Model C: } &\beta_1+\beta_2{\rm male}_j+\alpha_{{\rm a}[j]}^{\rm age}+\alpha_{{\rm r}[j]}^{\rm race}+\alpha_{{\rm t}[j]}^{\rm time}+\alpha_1{\rm ADI}_{{\rm s}[j]} +\alpha_2{\rm college}_{{\rm s}[j]}+\alpha_3{\rm employment}_{{\rm s}[j]} +\alpha_4{\rm income}_{{\rm s}[j]}+\alpha_5{\rm poverty}_{{\rm s}[j]} +\alpha_6{\rm urbanicity}_{{\rm s}[j]}.
% \end{align*}
% }

\begin{table}
\centering
\resizebox{\textwidth}{!}{
\begin{tabular}{lcccccccccccc}
Model & Male & Age & Race & Time & ADI & College & Employment & Income & Poverty & Urbanicity & Race$\times$College & ZIP-varying effect \\
\hline
A & $\checkmark$ & $\checkmark$ & $\checkmark$ & $\checkmark$ & $\checkmark$ & $\checkmark$ & $\checkmark$ & $\checkmark$ & $\checkmark$ & $\checkmark$ &  & $\checkmark$ \\
B  & $\checkmark$ & $\checkmark$ & $\checkmark$ & $\checkmark$ & $\checkmark$ & $\checkmark$ & $\checkmark$ & $\checkmark$ & $\checkmark$ & $\checkmark$ & $\checkmark$ & $\checkmark$ \\
C& $\checkmark$ & $\checkmark$ & $\checkmark$ & $\checkmark$ & $\checkmark$ & $\checkmark$ & $\checkmark$ & $\checkmark$ & $\checkmark$ & $\checkmark$ &  & 
\end{tabular}
}
\caption{Predictors included in each model. A checkmark indicates inclusion.}
\label{model:display}
\end{table}

\begin{table}
\centerline{
\begin{tabular}{ccc}
          &  elpd\_diff  & se\_diff\\
 \hline         
Model A    & 0   &    0  \\
Model B & $-2.05$   &    0.86   \\
Model C & $-4.75$    &   3.20 \\
\end{tabular}
}
\caption{Model comparisons with the leave-one-out cross-validation. The two columns show the expected log predictive density difference and its standard error, in each case comparing to Model A. The negative elpd\_diff values show that Model A has the best predictive performance. The se\_diff values support whether the improvement of Model A is substantial.}
\label{loo}
\end{table}

Table \ref{loo} gives the LOO-CV outputs on the model comparison. The difference, elpd\_diff, will be positive if the expected predictive accuracy for Model B or Model C is higher than that for Model A. The negative elpd\_diff values show that Model A has the best predictive performance. The se\_diff values support whether the improvement of Model A is substantial. A rule of thumb is to check whether the interval (elpd\_diff- 2$*$se\_diff,  elpd\_diff + 2$*$se\_diff) covers the value 0. Hence, we select Model A for inference.

\begin{figure}[htp]
\begin{tabular}{c}
\includegraphics[width=0.95\textwidth]{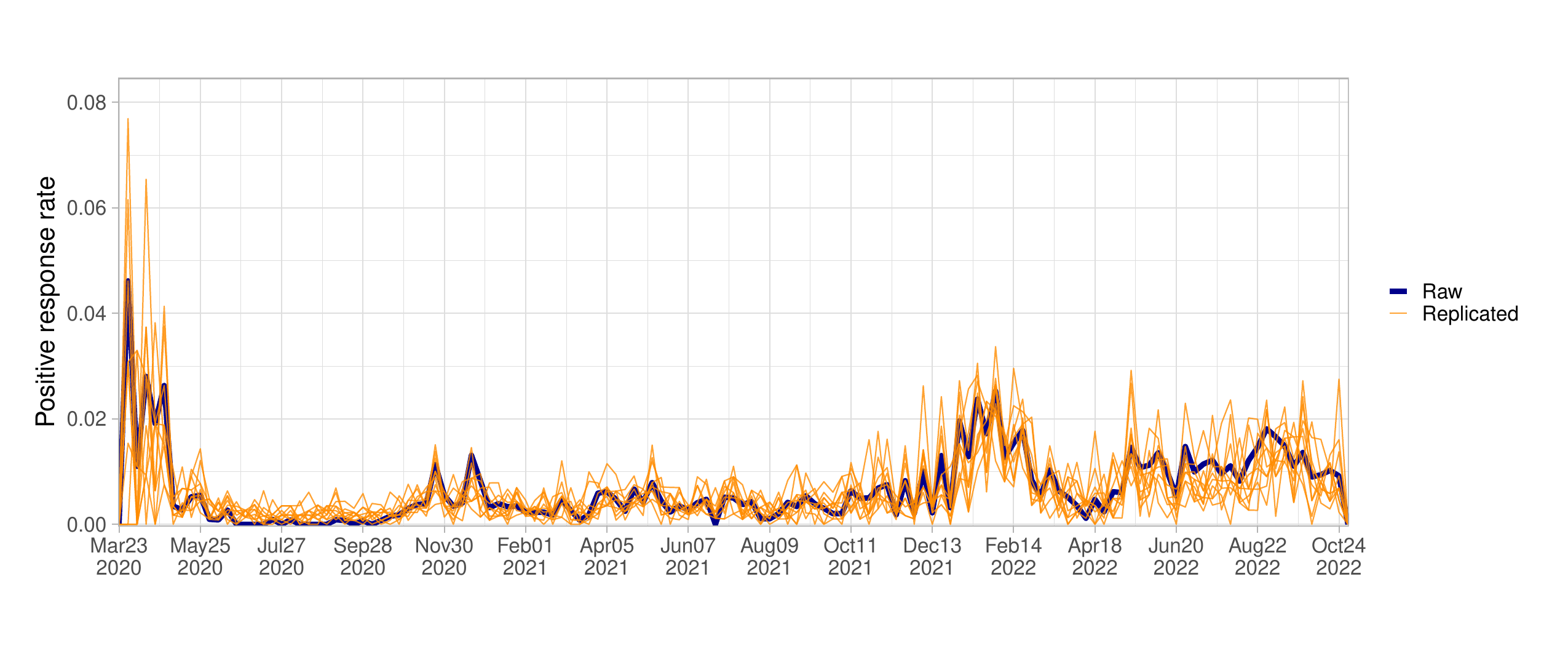} 
\end{tabular}
\caption{Posterior predictive check comparing replicated positive response rates generated from the estimated model to raw rates by week. The replicates are close to the raw data, showing that the model fits the data well.}
\label{ppc}
\end{figure}

If the model fits the observed data well and preserves the correlation structure, we expect the model to generate replicated data of the observations that mimic the raw values. The weekly replicates use the observed number of tests and estimated positivity based on models~\eqref{positivity} and \eqref{pi} corresponding to each week to generate synthetic counts of positive cases and then the synthetic positive response rates. The PPC in Figure~\ref{ppc} examines the weekly positivity and compares the raw values to 10 sets of replicates based on posterior predictive samples from Model A. When the number of tests is small, the generated replicates present large variability. Across time, the replicates are close to the observations, showing that the model fits the data well without red flags of failing to capture important structure. 

\subsection{Estimation results}
\label{results}

Based on the selected Model A, we present the estimated infection incidence over time for the target population and demographic and geographic subpopulations.

\begin{figure}[htp]
\begin{tabular}{c}
\includegraphics[width=0.95\textwidth]{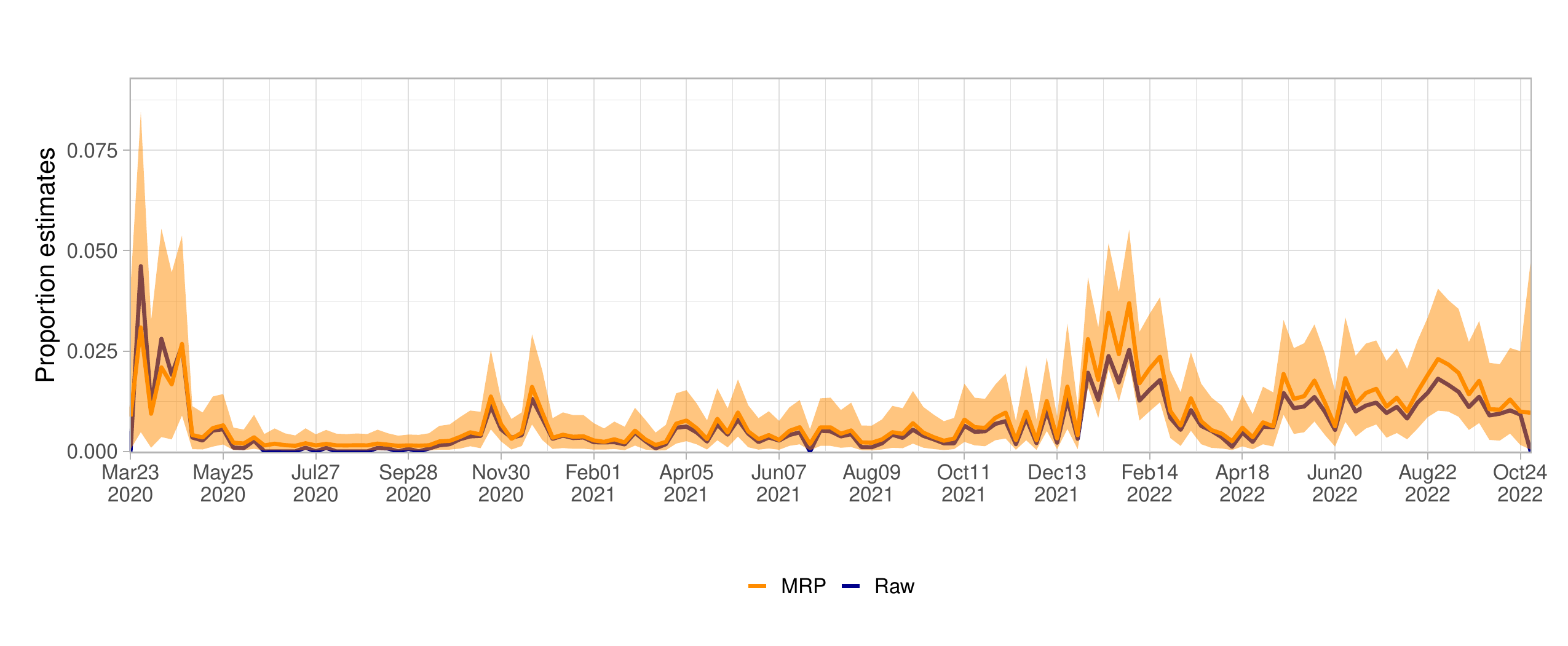}
\end{tabular}
\caption{Estimated weekly incidence in the community based on the multilevel regression and poststratification (MRP) in comparison with raw values. The shaded areas represent 95\% credible intervals. The MRP estimates match the sample demographics with those in the population and are generally higher than the raw positivity, mainly because of the test sensitivity.}
\label{all_est}
\end{figure}

Figure~\ref{all_est} shows the estimated viral transmission rate by week in the catchment area. We observe spikes in November 2020, January 2022, and August 2022. The MRP estimates are generally higher than the raw positivity, mainly because of the test sensitivity, where 70\% of infections are tested positive. MRP matches the sample demographics with those in the population. 

\begin{figure}[htp]
\begin{tabular}{c}
\includegraphics[width=0.95\textwidth]{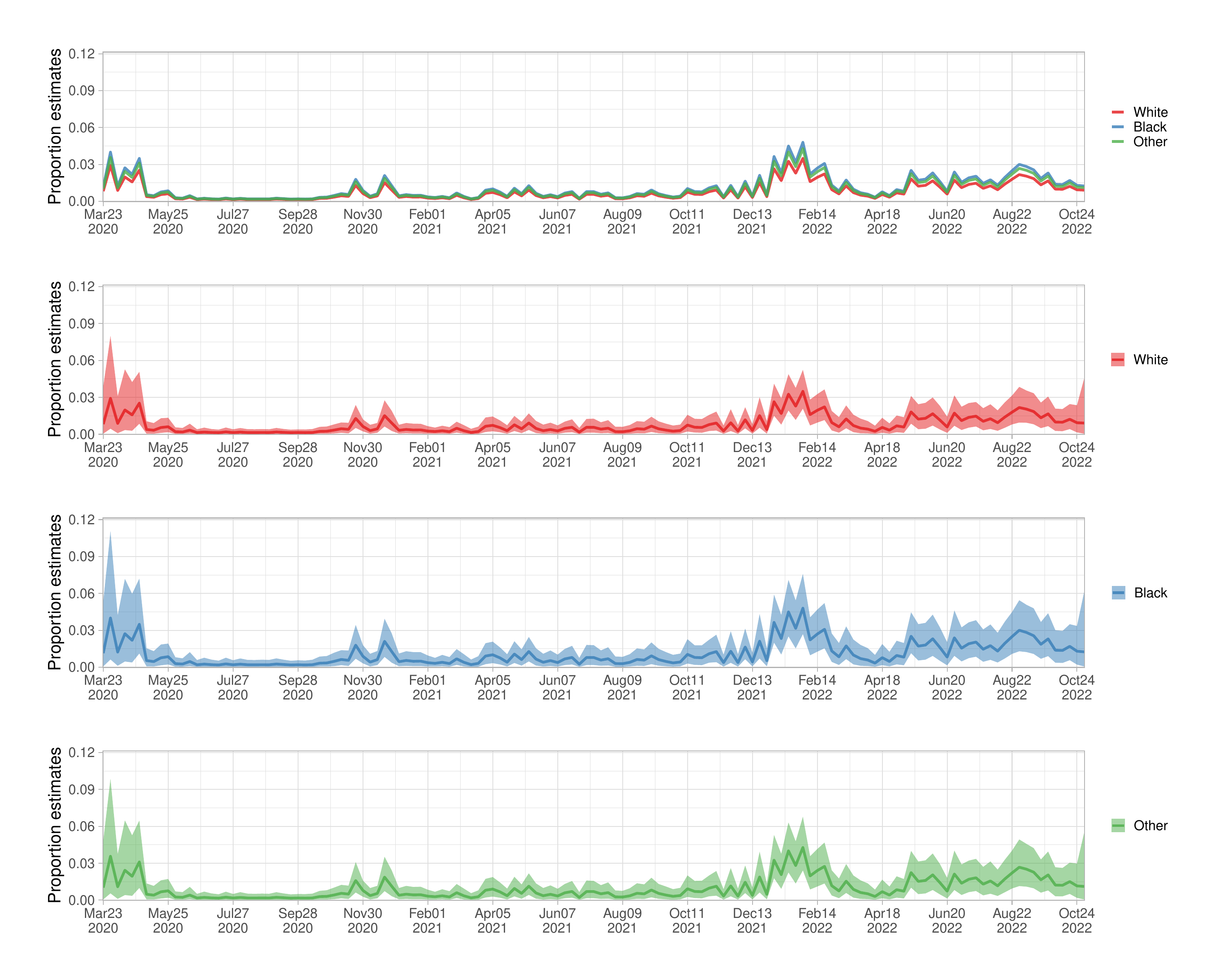} 
\end{tabular}
\caption{Estimated weekly incidence across racial groups based on the multilevel regression and poststratification. Whites tend to have lower infection rates than Black and other racial groups, even though most weekly differences are small. The shaded areas represent 95\% credible intervals.}
\label{r_est}
\end{figure}

MRP stabilizes small group estimates and adjusts for the sample discrepancy within each group. Figure~\ref{r_est} presents the estimated incidence for White, Black, and other race categories. Whites tend to have lower infection rates than Black and other racial groups, even though most weekly differences are small and not significant. The model does not include time trends varying across racial groups, i.e., without racial moderation effects. The estimated trends are close to paralleling with similar spike and flat periods. Examining racial differences by week, we expect the differences to be small because of weekly small numbers of tests. When we calculate the cumulative incidences through the study period, Whites are less likely to be infected than Black and other racial groups, which is consistent with the literature findings~\citep{Magesh_2021}. The same observation of trends applies to the sex and age group estimates, given in Figures~\ref{s_est} and \ref{a_est} of the Appendix.

\begin{figure}[htp]
\begin{subfigure}{0.32\textwidth}
    \centering
\includegraphics[width=0.99\textwidth]{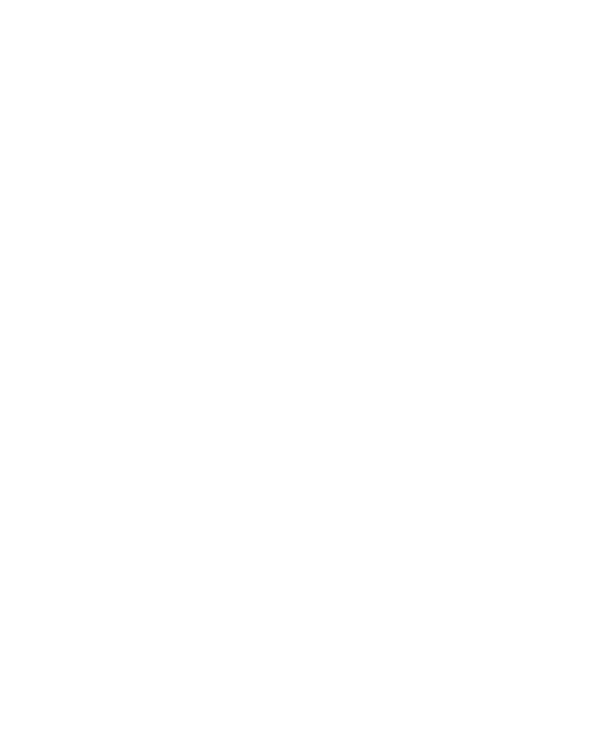} 
    \caption{Observed positivity}
\end{subfigure}
\begin{subfigure}{0.32\textwidth}
    \centering
\includegraphics[width=0.99\textwidth]{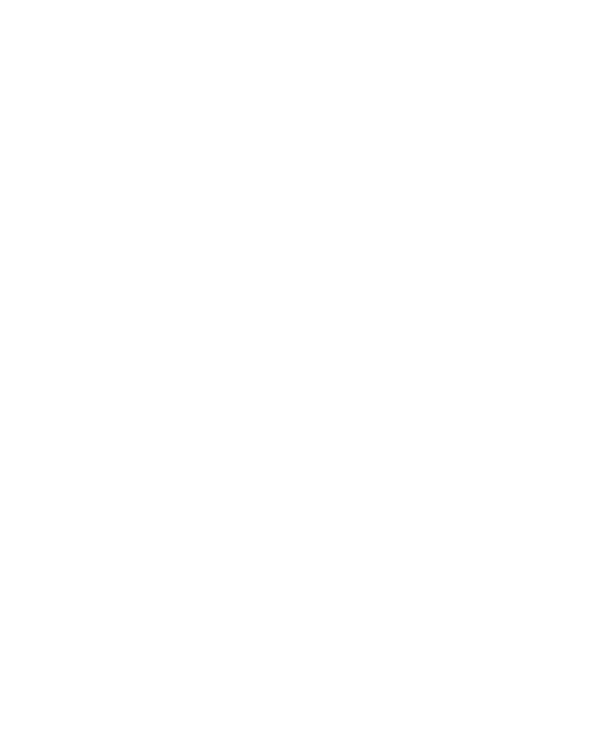} 
    \caption{Estimated incidence}
\end{subfigure}
\begin{subfigure}{0.32\textwidth}
    \centering
\includegraphics[width=0.99\textwidth]{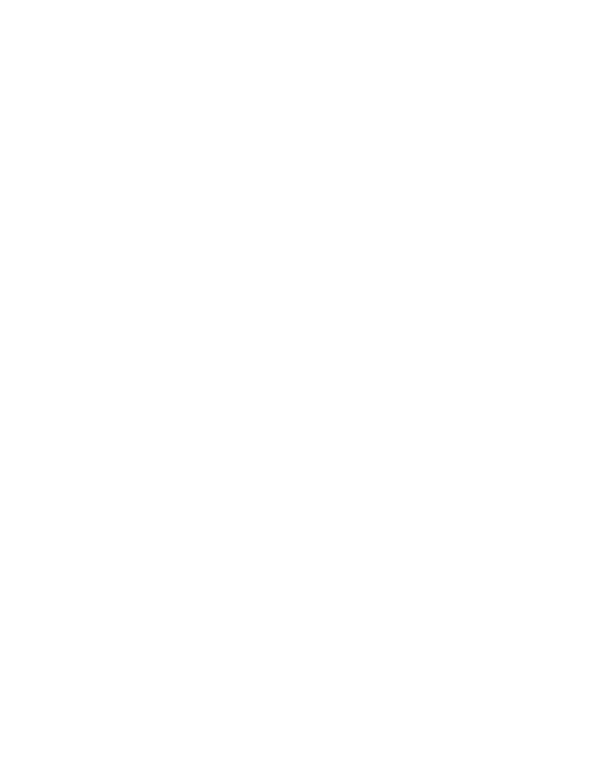}
    \caption{Standard errors}
\end{subfigure}\\
\caption{Observed and estimated county-level incidence with standard errors during the week of 01/31/2022--02/06/2022 in the catchment area. The model estimates reduce between-county variation, and their standard errors generally increase with the estimated incidence values. }
\label{c_est}
\end{figure}

We select one week that observes a spike of infection, 01/31/2022--02/06/2022, and present the county-level estimated incidence with standard error (SE) values in Figure~\ref{c_est}. The collected test records across counties are sparse, where 32 out of 94 counties do not have any tests and 28 counties have only one test during the selected week. The county with the largest number of tests (230) is where the health system is located. The observed positivity values are unreliable, and 52 out of 62 collected values are zero. Among the top five counties that report the largest numbers of tests and any positive cases during the studied week, the demographic distributions of tested patients are generally similar, with an over-representation of female, White, and older people compared to the ACS data. The MRP estimates are available for all 94 counties based on the predictions with the ACS data. The model fit summaries in the Appendix A show that the estimated coefficient of urbanicity (defined as the percentage of covered census tracts classified as urban, weighted by tract population) is $-0.10$ with the 95\% interval of $(-0.21, 0.02)$, which shows that the urban areas tend to have lower infection rates than the non-urban areas. The multilevel model smooths county-level incidence estimates with a range of 0.1\%--5.6\%, with a median value of 0.6\%. The SE values of the 94 county-wise incidence estimates are between 0.001 and 0.037, and the variation generally increases with the estimated incidence. 

\subsection{Validation and comparison}
\label{validation}

\begin{figure}[htp]
\begin{tabular}{c}
\includegraphics[width = 0.8\textwidth, height=2.5in]{plot/overall_est.pdf}\\
\includegraphics[width = 0.8\textwidth, height=2.5in]{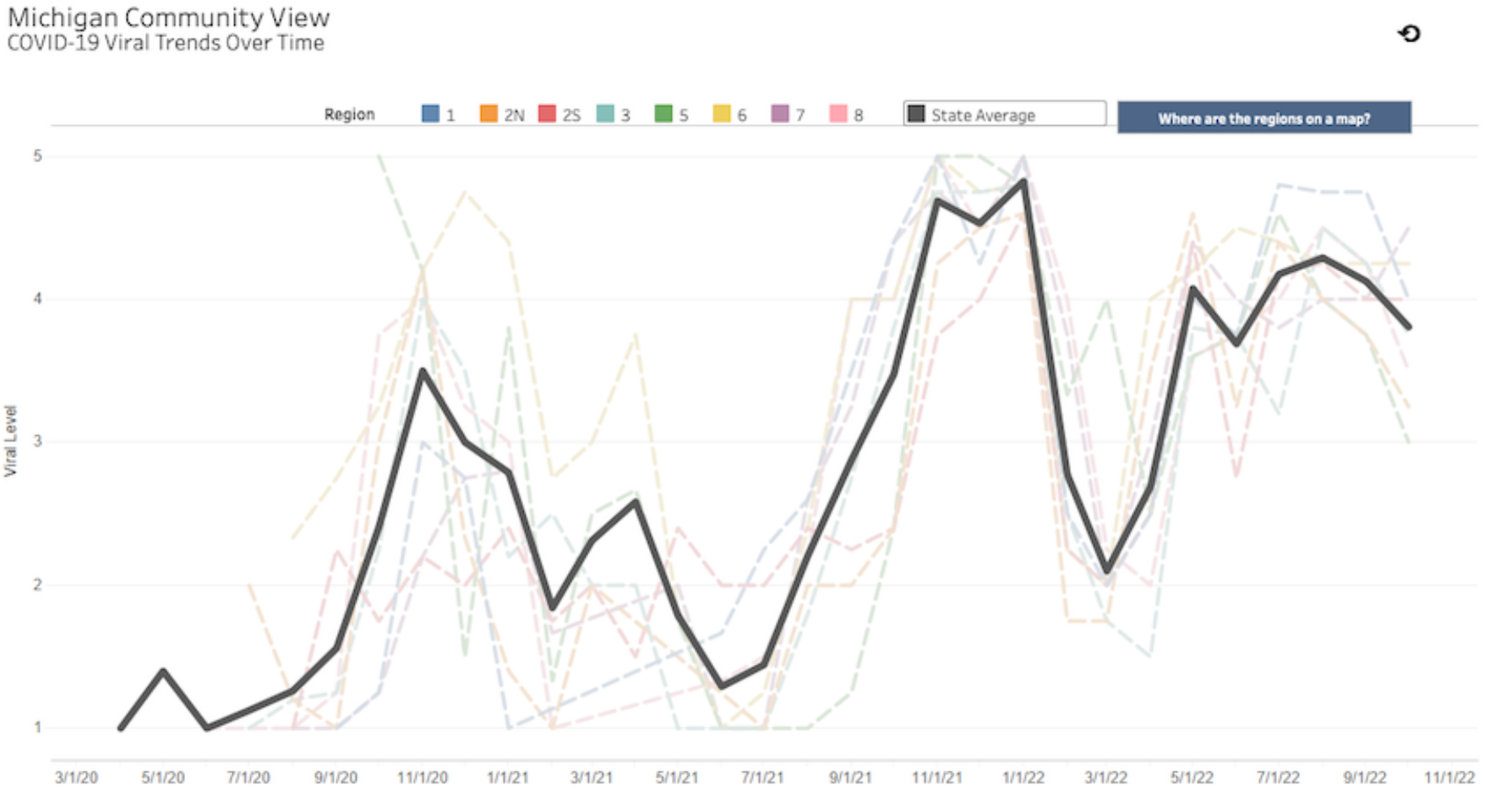}\\
\includegraphics[width = 0.99\textwidth, height=2.9in]{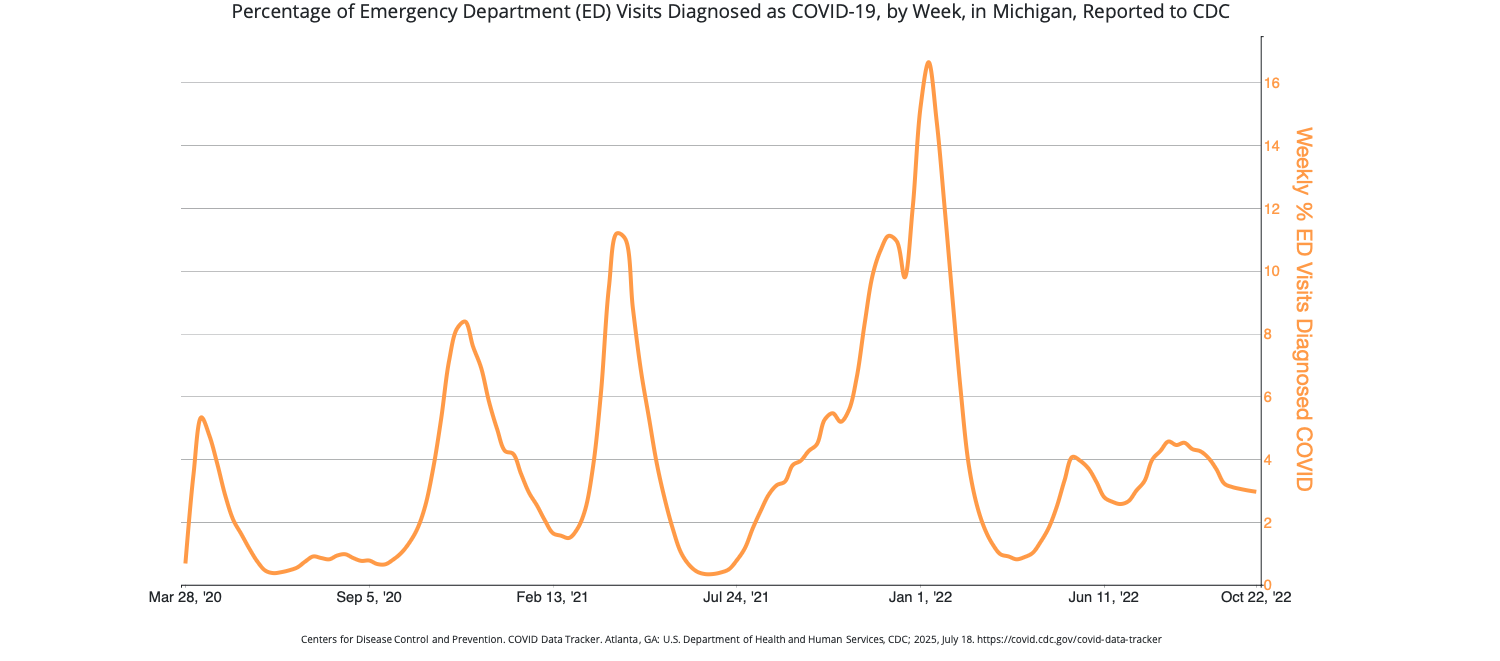}
\end{tabular}
\caption{Comparison of COVID-19 tracking trends estimated by applying multilevel regression and poststratification (MRP) to Michigan hospital test data, the Michigan wastewater monitored COVID-19 surveillance and weekly percentages of Emergency Department (ED) visits diagnosed as COVID-19 in Michigan, showing an alignment of transmission trends and early warning of clinical burdens.}
\label{mi-val}
\end{figure}

Our surveillance tool leverages routine hospital testing of asymptomatic patients to provide an early indicator of community disease presence, serving a similar function to wastewater monitoring for SARS-CoV-2 in public sewer systems. By tracking trends over time, both tools can detect increases in SARS-CoV-2 prevalence, thereby alerting health agencies to potential surges in cases and an increased clinical burden. We have compared our results to data from the Michigan Wastewater Dashboard for COVID-19 surveillance~\citep{MIwastewater} and to weekly percentages of Emergency Department (ED) visits diagnosed as COVID-19 in Michigan, as reported by the CDC COVID Data Tracker~\citep{CDCcovidtracker}. Figure~\ref{mi-val} demonstrates that our estimated trends between March 22, 2020, and October 24, 2022, closely align with wastewater-based surveillance for SARS-CoV-2 shed into Michigan’s public sewer systems. Both surveillance methods capture the spikes in November 2020 and January 2022, as well as the downward trend since August 2022. Notably, our surveillance approach can anticipate increases in ED visit numbers reported to the COVID Data Tracker by approximately one to two weeks. This validation supports findings previously reported in Indiana~\citep{mrp-covid21, mrp-covid22}.

Our results based on a representation adjustment of routine hospital test records serve a synthetic proxy for random sampling. When available, random-sample testing surveys provide valuable benchmarking data and should be leveraged to calibrate other data sources to ensure population representativeness~\citep{irons2021estimating,menachemi2020population}. However, increasing nonresponse rates in these surveys necessitate nonresponse bias adjustments~\citep{NRBA-cr2022, NRBA-long2022}. For example, \cite{yiannoutsos2021bayesian} have applied a similar method to MRP and adjusted for nonresponse bias in a randomized study of COVID-19 testing in Indiana, the response rate of which is 23.6\%. Notably, the trends in calibrated new infection numbers reported by~\cite{irons2021estimating} are consistent with those seen in the MRP-adjusted hospital test incidence monitoring~\citep{mrp-covid21} between March 2020 and March 2021 in Indiana, particularly regarding the capture of infection spikes.

\section{Discussion}
\label{discussion}

With generalizability as the goal, the MRP method adjusts for selection bias and stabilizes small group estimates. We extend MRP to both cross-sectional and longitudinal data and facilitate subgroup trend analyses over time with a user-friendly interface. A key contribution of our MRP interface is its capacity to standardize statistical workflows and facilitate reproducibility. By integrating all analytic steps---from data preprocessing and model fitting to diagnostics and result visualization---into a comprehensive pipeline, the platform makes it straightforward for others to replicate findings and validate results on new datasets. The interface is built using open-source tools (Stan, R, Shiny) and is publicly available, supporting transparency and broad accessibility. By making MRP methods widely available without substantial coding requirements, the interface empowers a wider range of users to implement advanced statistical methods and strengthens training in reproducible population health analytics. These advantages position the interface as a foundational resource for standardized use and methodological rigor in future epidemic surveillance and related scientific studies.

The interface tracks the COVID-19 epidemic and delivers substantive findings in time, which is demonstrated using the Michigan health system data. The findings based on the MRP workflow have implications for public health monitoring and policy decision-making. By correcting sample selection bias and generating stable estimates for small demographic and geographic subgroups, the approach can identify populations at higher risk of infection or with lower access to healthcare resources. Health agencies can use subgroup-specific incidence estimates to allocate testing resources and direct targeted communication to communities with greater vulnerability. Tracking epidemiological trends using routine hospital data provides actionable guidance for predicting surges, anticipating clinical burden and responding swiftly to local outbreaks. Furthermore, the analytic pipelines support transparency and comparability in health reporting, which is vital for evidence-based policymaking and resource allocation.

Our approach has several key assumptions. First, we assume sample selection is ignorable, conditional on adjusted demographics and geography. The only factor determining the sample inclusion is the selection for elective surgical procedures. Elective surgery patients may differ from the broader community in unmeasured ways, such as healthcare access or overall health status. However, we need high-quality population data on healthcare use and health measures to adjust these potentially confounding factors. Second, we assume a constant ratio of asymptomatic to symptomatic SARS-CoV-2 infections within any demographic and geographic stratum. We use the estimated incidence based on asymptomatic test records to track the infection trend, but not the magnitude. The ratios may change values with new viral variants and immunity levels that are naturally-acquired or vaccine-induced. We have conducted sensitivity analyses by including the time indicators when the new viral was first detected in Michigan in the model and found that the estimated trends are similar, though the incidence rates have slightly shifted. We have also applied the MRP adjustment to viral IgG testing data of the same group of asymptomatic patients and validated the method using verified clinical metrics of viral and symptomatic disease incidence to show the expected biological correlation of these entities with the timing, rate, and magnitude of seroprevalence~\citep{mrp-covid22}. Third, the model-based adjustment is subject to model misspecification. We use a Bayesian binomial model, and our subgroup estimates are robust across different outcome models. It is possible that the model fails to capture some data structure. We suggest users conduct thorough model diagnostics, such as the PPC and LOO-CV in our paper, and result validation. Additionally, the interface’s current focus is infection incidence estimation by subgroup, but it is extensible to other epidemiological parameters (e.g., effective reproduction number, infection fatality ratio), which would require integrating individual-level test data with aggregate case and mortality data using hierarchical Bayesian frameworks. Despite these limitations, the flexibility of the MRP approach allows its use with a variety of sampling methods and data sources, supporting broader applicability beyond the specific settings tested so far. Though post-epidemic testing is currently paused, the interface has broad applicability for other disease monitoring and data analyses with population representation. 

With respect to our current sampling method for COVID-19 viral tracking, in accordance with accepted American Society of Anesthesiology standards, all preoperative patients in the hospital system are subjected to surgical risk evaluation. Hence, routine hospital testing is already implemented uniformly across the U.S., increasing the operational feasibility of our proposed surveillance system. Previous work used data from a community hospital in Indiana~\citep{mrp-covid21}, and the results show differences between the states. As shown in our county-level estimates in Michigan, geographic variation can be substantial. Expanding to data from more states will further enhance national generalizability. Empowering users of varying backgrounds to analyze their own localized data effectively, the workflow can be directly adopted by local, state, and national health departments to ensure methodological consistency. 

Originally developed in response to COVID-19, the MRP interface provides a foundation for broader epidemic surveillance and diverse applications in health and social science research. It accommodates time-varying and cross-sectional data, continuous and binary outcomes, and supports subgroup analyses across demographic and geographic domains. Users can specify models, priors, and poststratification data, and analyze probability sample surveys, non-probability samples, and multiple data sources. Future enhancements will address complex spatiotemporal structures (e.g., integrating real-time individual and aggregate testing data with autoregressive structures across transmission neighborhoods), custom prior distributions, and poststratification with incomplete population data. The MRP workflow is broadly applicable across a range of public health surveillance contexts. For example, recent measles and pertussis outbreaks in various parts of the U.S. underscore the need for flexible, reliable surveillance systems that can rapidly estimate disease prevalence and identify at-risk populations \citep{cdc-measles, cdc-pertussis}. By facilitating subgroup estimation and adjusting for selection bias, our MRP interface could be readily adopted for outbreak response and ongoing surveillance of many infectious diseases, where sample representativeness and rapid trend analysis are critical. The potential to use routinely collected hospital data, school outbreaks, or sentinel surveillance and combine them with population-level information makes our framework adaptable beyond COVID-19, reinforcing its generalizability for future epidemic preparedness and monitoring of emerging health threats.

In summary, the MRP interface advances statistical methods for epidemic surveillance by combining methodological rigor, computational accessibility, and practical impact. As public health agencies move toward more data-driven and adaptive response models, tools that enable granular, timely, and reproducible estimation will become increasingly essential. By establishing a foundation for standardized population health analytics, our approach aims to support ongoing efforts in epidemic preparedness, resource optimization, and evidence-based policy.

\section*{Acknowledgments}
This work is supported by the National Institutes of Health through grant U01MD017867. 

\section*{Data Availability}

The patient records of the COVID-19 tests are confidential and require data use agreement in compliance with HIPAA privacy regulations for access. The interface includes synthetic example datasets for replication and demonstration purposes. 

\appendix

The appendix provides supplemental analysis results. 

\section{Summary of geographic characteristics}

\begin{figure}[htp]
\begin{tabular}{cc}
\includegraphics[width=0.475\textwidth, height=2.9in]{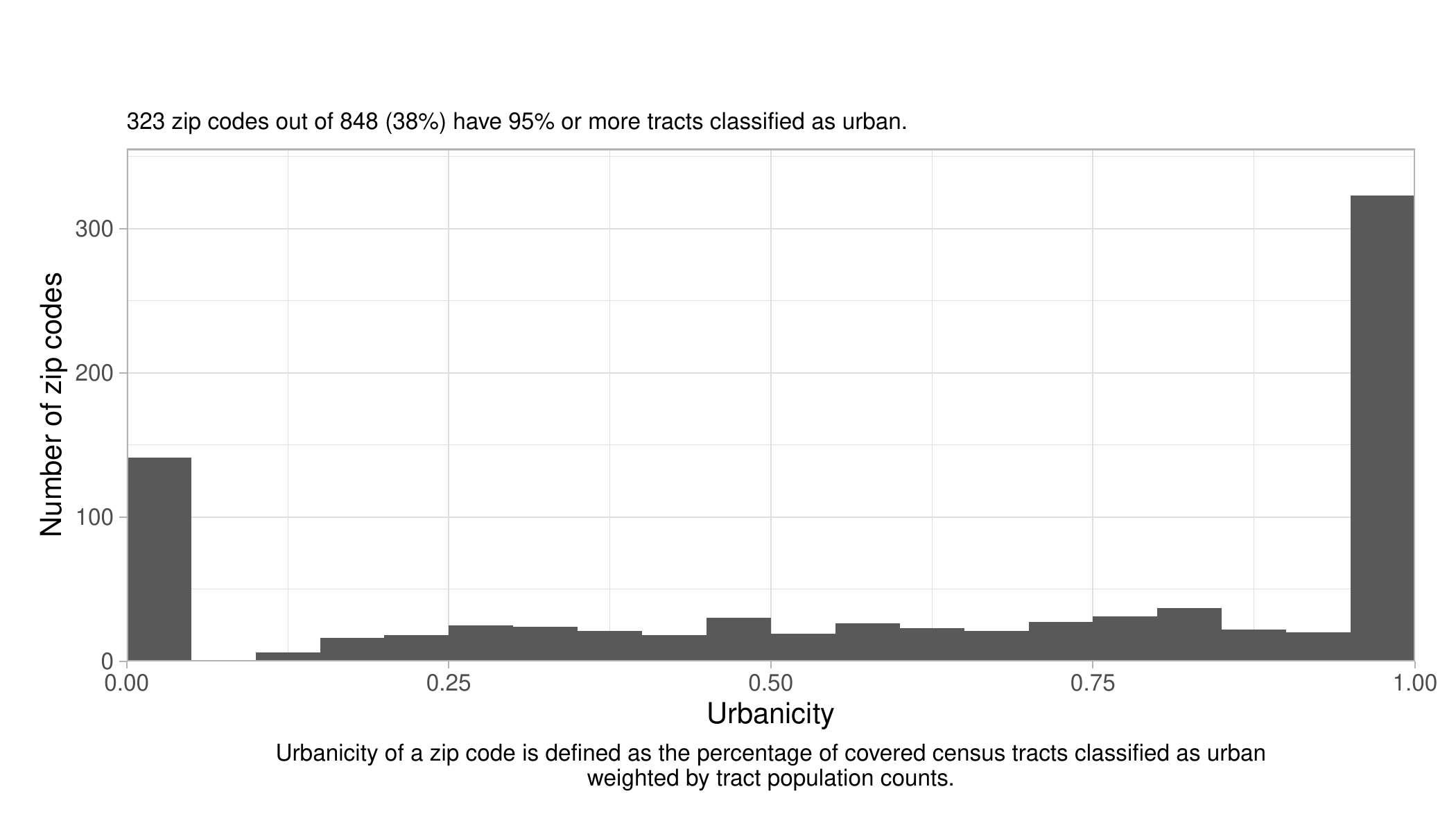} & 
\includegraphics[width=0.475\textwidth, height=2.9in]{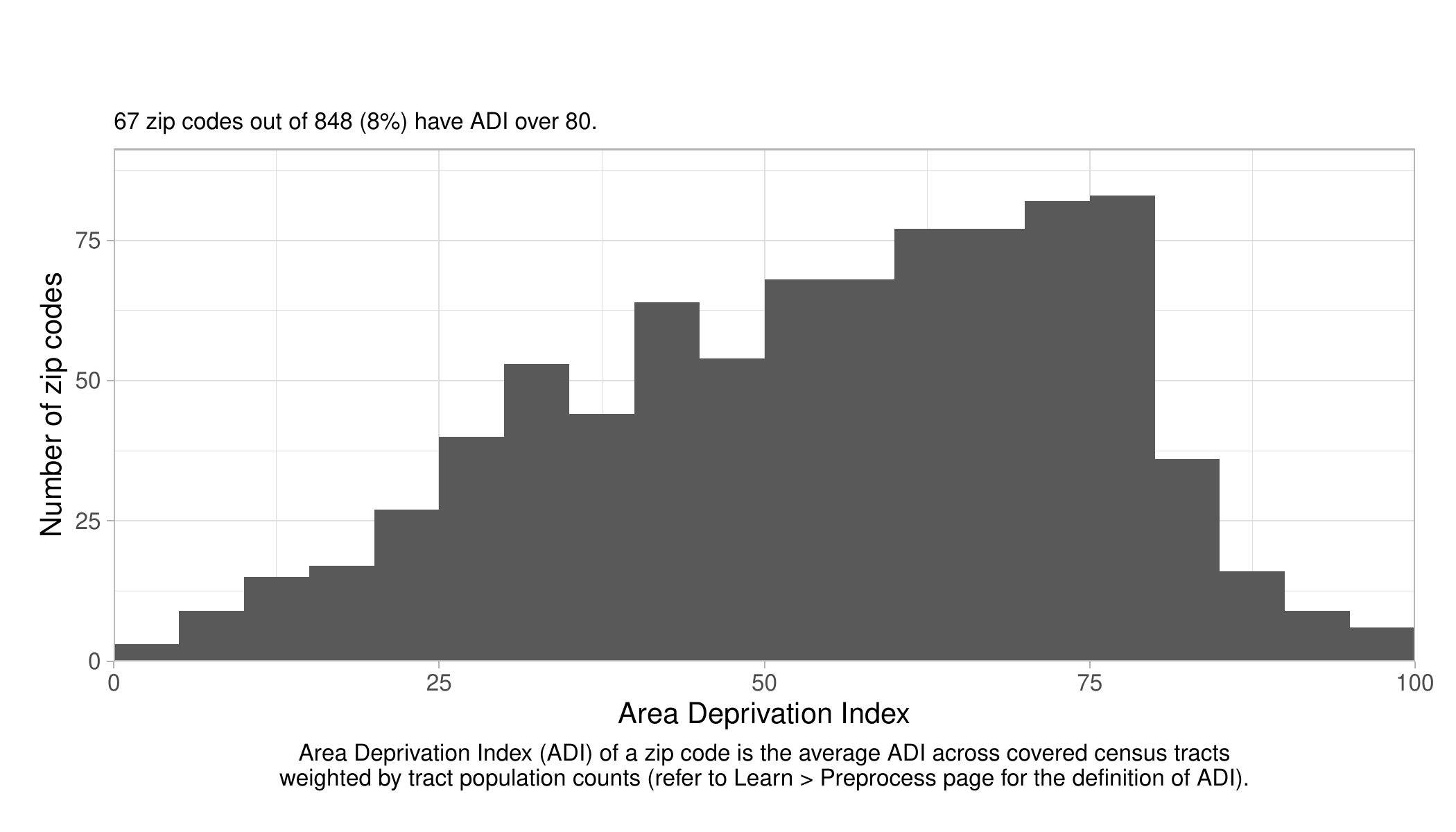} \\
\includegraphics[width=0.475\textwidth, height=2.9in]{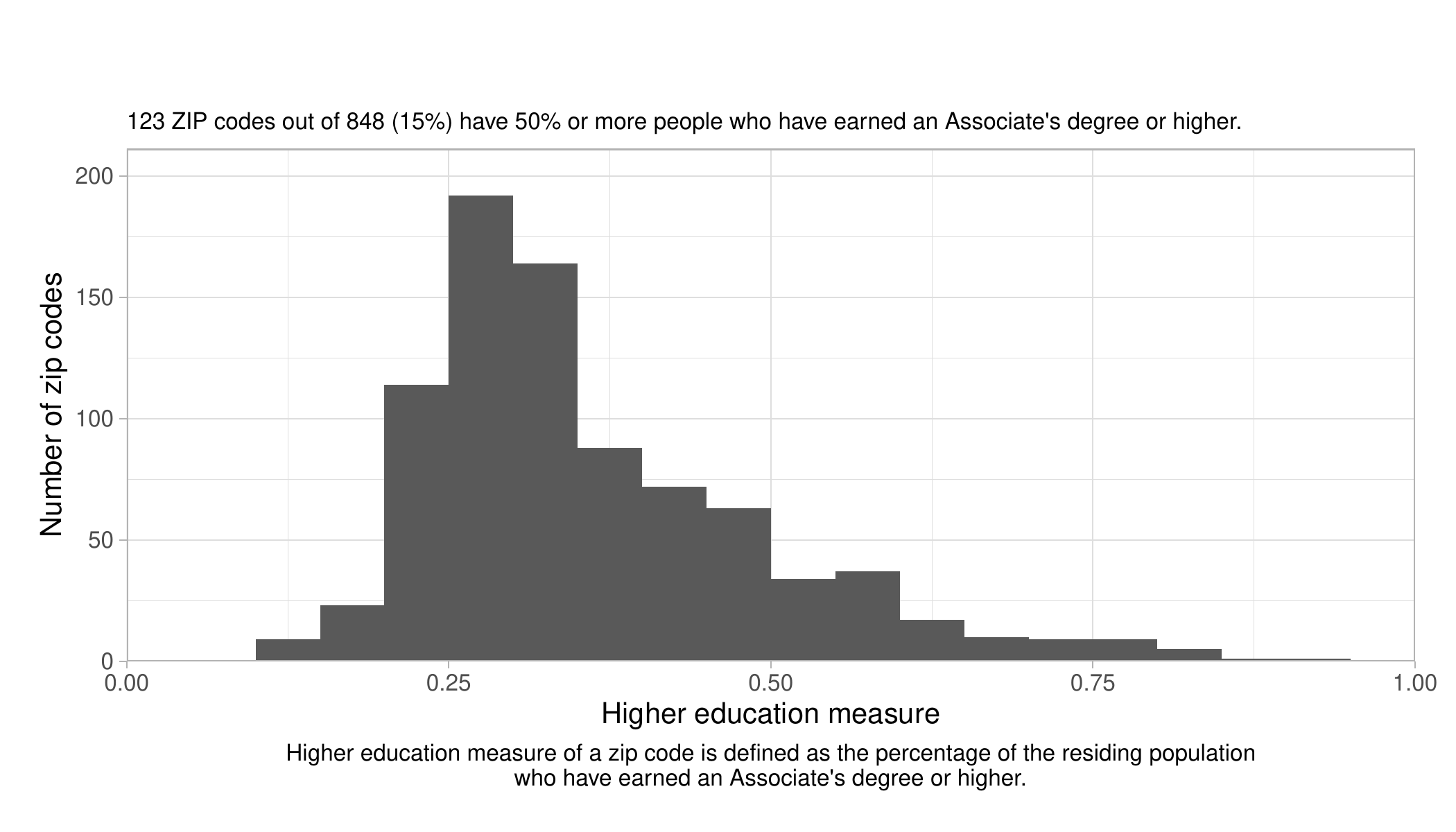} & 
\includegraphics[width=0.475\textwidth, height=2.9in]{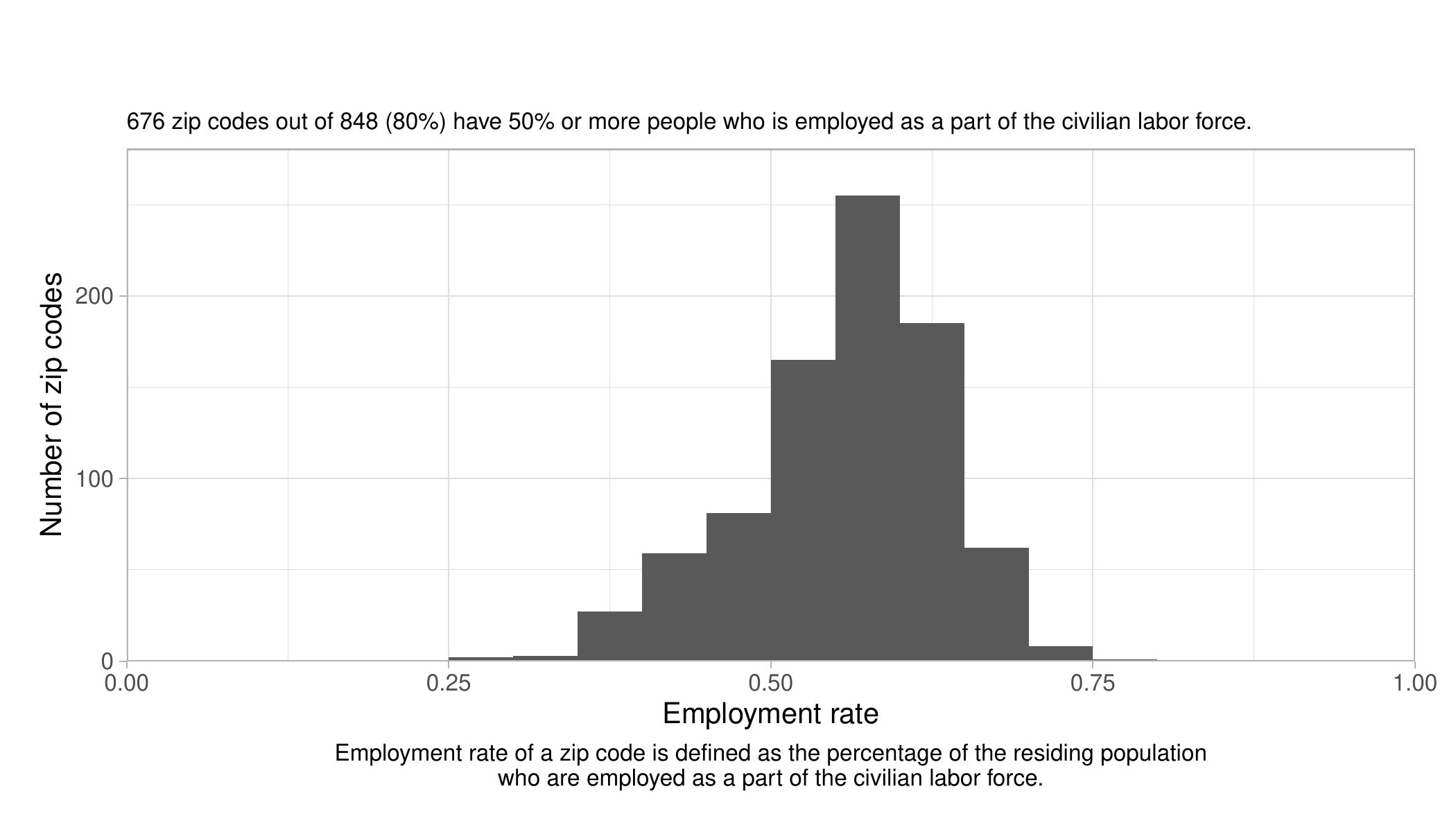} \\
\includegraphics[width=0.475\textwidth, height=2.9in]{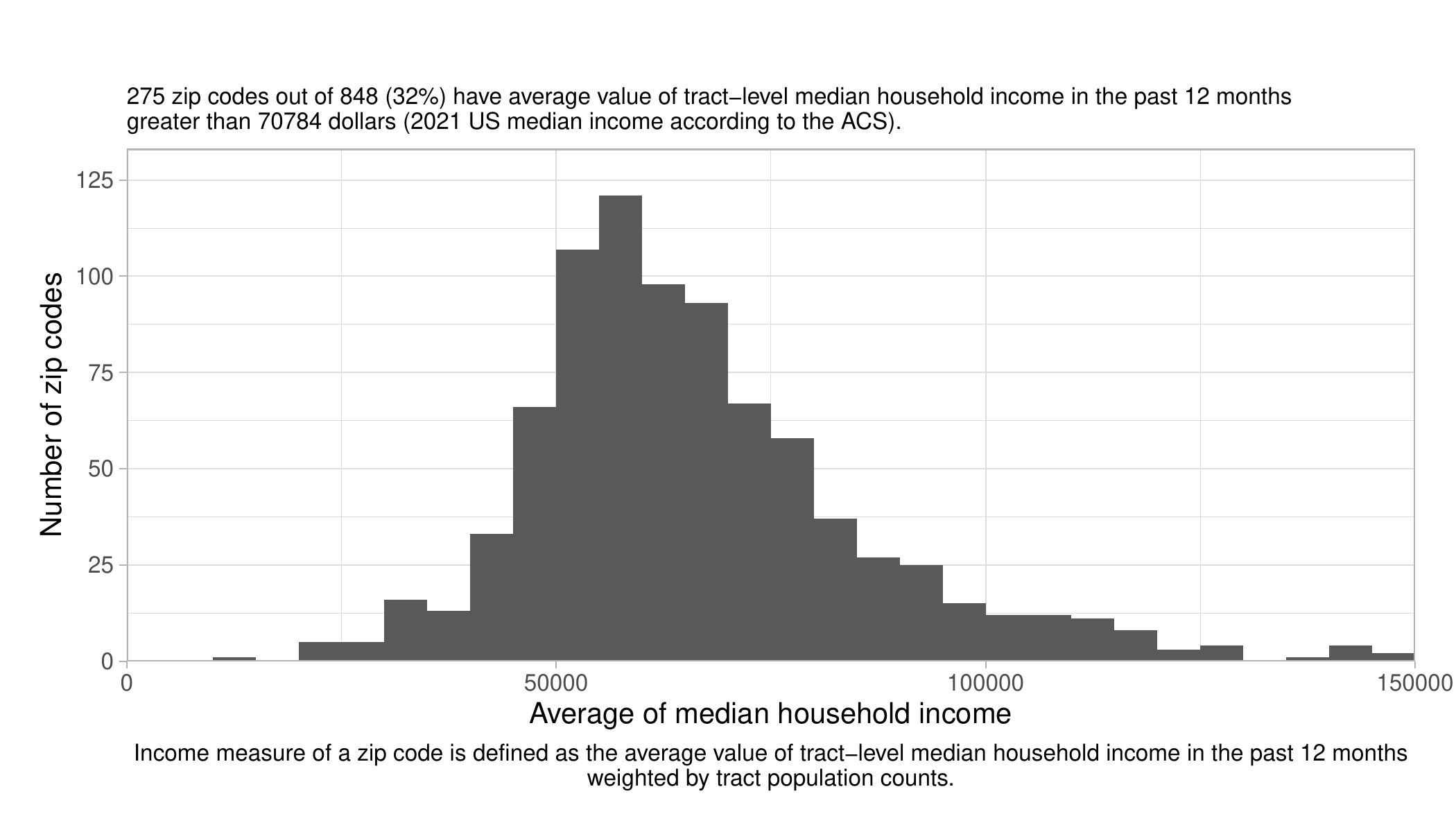} & 
\includegraphics[width=0.475\textwidth, height=2.9in]{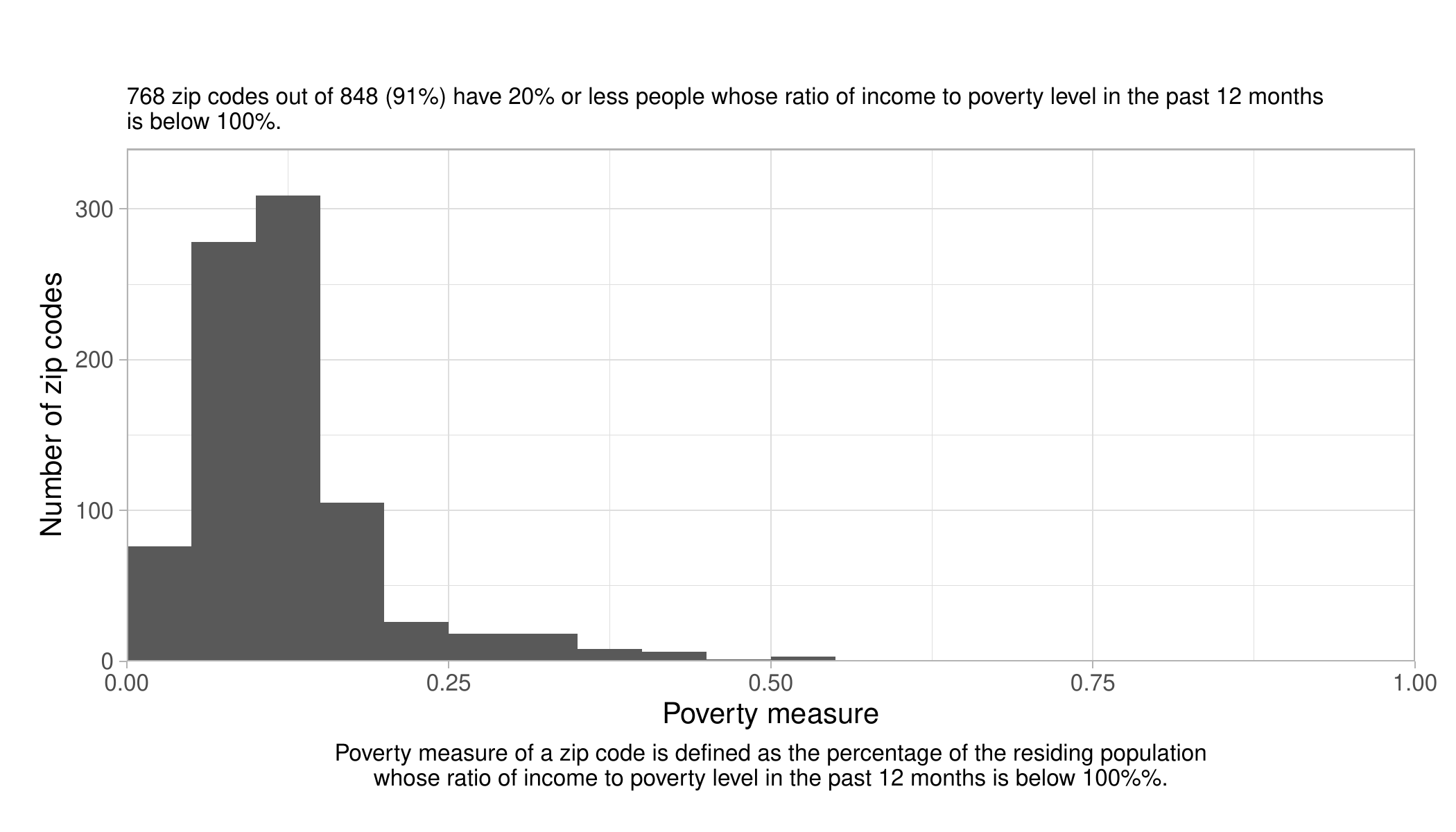} 
\end{tabular}
\caption{Distributions of geographic characteristics based on the linked American Community Survey data in the catchment area.}
\label{ZIP}
\end{figure}

Table~\ref{ZIP} gives the distributions of geographic characteristics.

\section{Model fitting results}
\label{fit_print}
The Stan fit summaries of Model A are displayed in Table~\ref{Stan-fit}.

A binomial model with a logit function of the positive response rate. Samples are generated using 4 chains with 2500 post-warmup iterations each. 

%\textbf{Non-varying Effects}
\begin{table}[htp]
\resizebox{\ifdim\width>\linewidth\linewidth\else\width\fi}{!}{
\begin{tabular}{lrrrrrrr}
\multicolumn{4} {l} {\bf Non-varying Effects}&&&&\\
& Estimate & Est.Error & l-95\% & u-95\% & R-hat & Bulk\_ESS & Tail\_ESS\\
\midrule
Intercept & -5.45 & 0.45 & -6.37 & -4.50 & 1.00 & 1629 & 1498\\
sex.male & 0.19 &  0.10 & 0.00 & 0.38 & 1.00 & 10062 & 5779\\
urbanicity & -0.10 & 0.06 & -0.21 & 0.02 & 1.00 & 6723 & 5479\\
college & -0.16 & 0.12 & -0.40 & 0.08 & 1.00 & 3681 & 5508\\
employment & 0.03 & 0.08 & -0.13 & 0.19 & 1.00 & 6066 & 6221\\
poverty & -0.07 & 0.10 & -0.27 & 0.12 & 1.00 & 4295 & 5208\\
income & -0.11 & 0.15 & -0.40 & 0.18 & 1.00 & 4385 & 4825\\
ADI & 0.07 & 0.10  & -0.13 & 0.28 & 1.00 & 5272 & 1902\\
\bottomrule
% \end{tabular}}
% \end{table}

% \textbf{Standard Deviation of Varying Effects}

% \begin{table}[htp]
% \label{county-dem}
% \resizebox{\ifdim\width>\linewidth\linewidth\else\width\fi}{!}{
% \begin{tabular}{lrrrrrrr}
\multicolumn{4} {l} {\bf Standard Deviations of Varying Effects}&&&&\\
% \toprule
 & Estimate & Est.Error & l-95\% & u-95\% & R-hat & Bulk\_ESS & Tail\_ESS\\
% \midrule
race (intercept) & 0.45 & 0.43 & 0.04 & 1.72 & 1.00 & 1941 & 1404\\
age (intercept) &  0.63 & 0.38 & 0.24 & 1.64 & 1.00 & 1843 & 3246\\
time (intercept) & 1.13 & 0.12 & 0.92 & 1.37 & 1.00 & 1882 & 4113\\
ZIP (intercept) & 0.36 & 0.11 & 0.11 & 0.57 & 1.01 & 989 & 596
\end{tabular}}
\caption{Model A fit summaries.}
\label{Stan-fit}
\end{table}

\section{Model estimates}

Figures~\ref{s_est} and \ref{a_est} present the weekly incidence estimates by sex and age groups, respectively.

\begin{figure}[htp]
\begin{tabular}{c}
\includegraphics[width=0.95\textwidth]{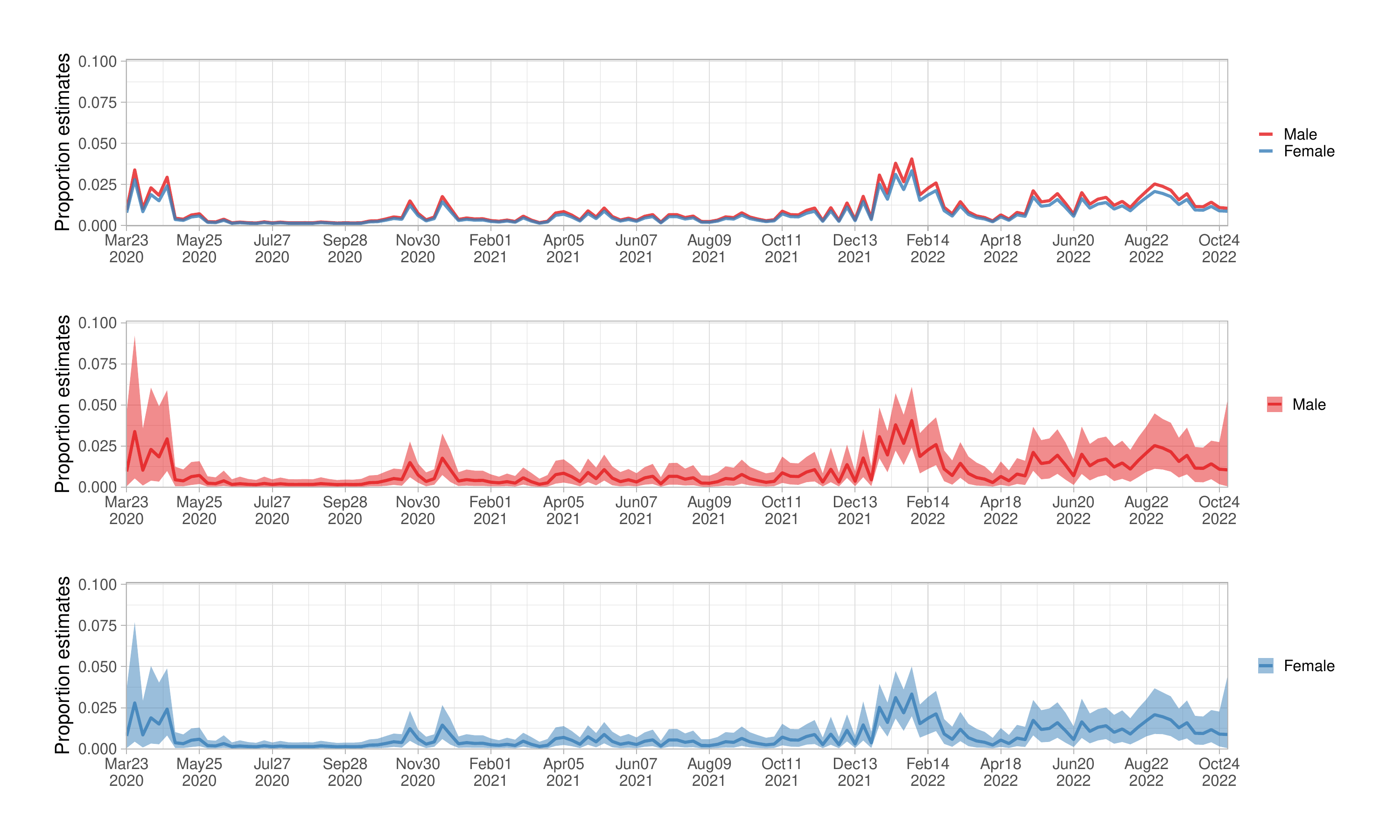} 
\end{tabular}
\caption{Estimated weekly incidence by sex based on the multilevel regression and poststratification. Females tend to have lower infection rates than males with small weekly differences. The shaded areas represent 95\% credible intervals.}
\label{s_est}
\end{figure}

\begin{figure}[htp]
\begin{tabular}{c}
\includegraphics[width=0.95\textwidth]{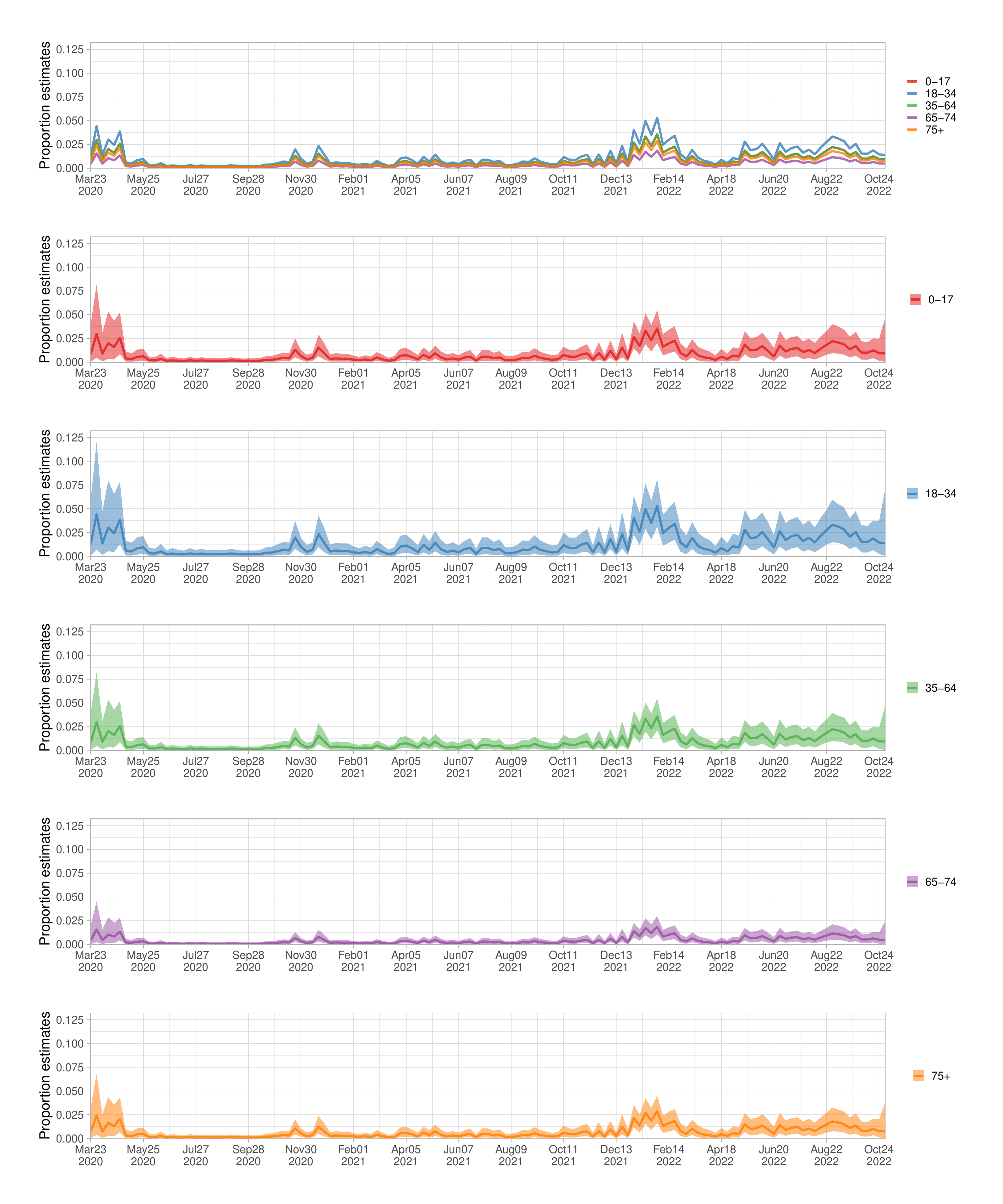} 
\end{tabular}
\caption{Estimated weekly incidence by age group based on the multilevel regression and poststratification. Young adults tend to have lower infection rates than elders with small differences during most weeks. The shaded areas represent 95\% credible intervals.}
\label{a_est}
\end{figure}

\bibliographystyle{chicago}
\bibliography{interface.bib}

\end{document}